\documentclass[twocolumn]{aastex631}

\newcommand{\degree}{${^\circ}$}
\newcommand{\reg}{\textsuperscript{\tiny\textregistered}}

\usepackage{soul}
\usepackage{xcolor}

\usepackage{comment}
\usepackage{wrapfig}
\usepackage{tabularx}
\usepackage{parskip}
\usepackage{textcomp}
\usepackage{subfigure}
\usepackage{placeins}

\begin{document}

\title{EDGES-3: Instrument Design and Commissioning}


\author[0000-0003-0267-8432]{Rigel C. Cappallo}
\affiliation{Massachusetts Institute of Technology, Haystack Observatory, Westford, MA 01886, USA}

\author[0000-0003-1941-7458]{Alan E. E. Rogers}
\affiliation{Massachusetts Institute of Technology, Haystack Observatory, Westford, MA 01886, USA}

\author[0000-0003-4062-4654]{Colin J. Lonsdale}
\affiliation{Massachusetts Institute of Technology, Haystack Observatory, Westford, MA 01886, USA}

\author[0000-0002-8475-2036]{Judd D. Bowman}
\affiliation{School of Earth and Space Exploration, Arizona State University, Tempe, AZ 85287, USA}

\author[0000-0002-9290-0764]{John P. Barrett}
\affiliation{Massachusetts Institute of Technology, Haystack Observatory, Westford, MA 01886, USA}

\author[0000-0003-3059-3823]{Steven G. Murray}
\affiliation{Scuola Normale Superiore, Piazza dei Cavalieri 7, 56126 Pisa, Italy}

\author[0000-0003-2560-8023]{Nivedita Mahesh}
\affiliation{Department of Astronomy, California Institute of Technology, Pasadena, CA 91125, USA}

\author[0000-0002-2871-0413]{Peter Sims}
\affiliation{School of Earth and Space Exploration, Arizona State University, Tempe, AZ 85287, USA}
\affiliation{Astrophysics Group, Cavendish Laboratory, J. J. Thompson Avenue, Cambridge CB3 0HE, UK}
\affiliation{Kavli Institute for Cosmology, Madingley Road, Cambridge CB3 0HE, UK}

\author[0000-0002-6611-2668]{Akshatha K. Vydula}
\affiliation{School of Earth and Space Exploration, Arizona State University, Tempe, AZ 85287, USA}

\author[0000-0002-3287-2327]{Raul A. Monsalve}
\affiliation{Space Sciences Laboratory, University of California, Berkeley, CA 94720, USA}
\affiliation{School of Earth and Space Exploration, Arizona State University, Tempe, AZ 85287, USA}
\affiliation{Departamento de Ingenier\'ia El\'ectrica, Universidad Cat\'olica de la Sant\'isima Concepci\'on, Alonso de Ribera 2850, Concepci\'on, Chile}

\author[0000-0003-2496-0383]{Christopher J. Eckert}
\affiliation{Massachusetts Institute of Technology, Haystack Observatory, Westford, MA 01886, USA}

\author[0000-0002-8489-0885]{Parker Steen}
\affiliation{Massachusetts Institute of Technology, Haystack Observatory, Westford, MA 01886, USA}

\author{Kenneth M. Wilson}
\affiliation{Massachusetts Institute of Technology, Haystack Observatory, Westford, MA 01886, USA}

\begin{abstract}

EDGES-3 is the third iteration of the EDGES experiment, designed to measure the predicted global absorption feature in the radio spectrum produced by neutral hydrogen gas at cosmic dawn, a critical observation determining when and how the first stars populated the universe.  The EDGES-3 instrument has been redesigned to include both the analog and digital electronics within the antenna, allowing for in-situ calibration and removal of the lossy balun found in EDGES-2.  EDGES-3 has been on multiple deployments in the past 4 years; to Oregon, Devon Island, Adak Island, and is currently installed and taking data in the outback of Western Australia.  This paper provides an accounting of the challenges inherent in the detection of the global, cosmological 21-cm signal, the strategies EDGES employs to mitigate each of these challenges, a description of the instrument, and a report on the Western Australia deployment along with observational data.

\end{abstract}



\section{Introduction} \label{sec:intro}

The cosmologically redshifted radio emission from the hyperfine spin-flip transition of neutral hydrogen is a uniquely powerful probe for exploring the cosmic Dark Ages and subsequent Epoch of Reionization (EoR), when the first stars and galaxies reionized the neutral hydrogen in the intergalactic medium (IGM). In observations of the signal at these cosmic distances, the 21-cm wavelength at emission has been redshifted into the very high frequency (VHF) radio band at tens to hundreds of megahertz (MHz). The hydrogen 21-cm spin temperature quantifies the relative number densities of atoms in the two hyperfine states. 21-cm cosmology experiments aim to measure the contrast between the 21-cm spin temperature and the radio background temperature, and both absorption and emission signatures can be used to trace evolving physical conditions as a function of cosmic time. Theoretically, such signatures are expected to originate from the coupling of the hydrogen 21-cm spin temperature to the IGM via the Wouthuysen-Field effect \citep{Wouthuysen1952}~thus making them a unique probe of inhomogeneous cooling, heating, and ionization influences in the early universe.  In addition, the spatially averaged spin temperature of the hydrogen gas is expected to decouple and diverge from the Cosmic Microwave Background (CMB) temperature, producing a frequency-dependent but direction-independent absorption and emission signature.  The redshifted 21~cm signal has been modeled to be in absorption at early times ($z\gtrsim12$), and emission at later times ($z\lesssim10$, \citealt[][and ref. therein]{Furlanetto2016}).  

Observing this redshifted 21-cm signal is hindered by bright Galactic foregrounds from synchrotron and free-free emission that are $10^4$ to $10^6$ times stronger than the expected signal. These foregrounds have substantial angular structure \citep{deOliveiraCosta2008}, but relatively little spectral structure, following an approximately power-law form \citep{Mozdzen2017, Mozdzen2019}.  Differences in spectral properties between the 21-cm signal and the foregrounds should enable separation of the two components.  Due to the large dynamic range, however, the foregrounds drive instrument design and require exquisite calibration and modeling of the instrument response during analysis.  Other foregrounds, such as Galactic radio recombination lines that do not follow the smooth synchrotron power-law, do not pose a significant challenge to 21-cm signal detection because they are narrow and easily excised in the broadband 21-cm observations \citep{Vydula2024}.  Additionally, ionospheric scintillation of strong galactic radio sources as well as terrestrial radio frequency interference (RFI) can further hamper a successful global 21-cm signal detection, requiring accurate data excision on time scales on the order of minutes.

Multiple experiments have been developed to target the redshifted 21-cm signal while mitigating the impact of bright foregrounds.  The experiments are divided into two categories: interferometers that measure spatial fluctuations of the signal, and small individual-antenna experiments that measure the all-sky average, or ``global'' 21-cm signal.  Building on early efforts by GMRT \citep{Paciga2013}, purpose-designed interferometers have since come online, including HERA \citep{Deboer2017, Berkhout2024}, LOFAR \citep{VanHaarlem2013}, OVRO-LWA \citep{Eastwood2019}, PAPER \citep{Parsons2010}, and the MWA \citep{Tingay2013}.  Over the last decade, these experiments have achieved increasingly stringent upper limits on the 21-cm power spectrum over redshifts $6<z<25$ \citep{Gehlot2019, Kolopanis2019, Mertens2020, Rahimi2021, HERA2023, Munshi2024} and developed foreground mitigation strategies \citep{Parsons2016, Chege2022, Kim2023}, sophisticated analysis methods \citep{Eastwood2018, Edler2021, Tan2021, Gorce2023, Chen2025}, techniques to better model and characterize instrument properties \citep{Jacobs2017, Josaitis2022, Rath2024}, and created sky maps and catalogs of the foregrounds \citep{HurleyWalker2017, Intema2017, Dowell2018, Shimwell2022}.  

Recent global 21-cm experiments include EDGES, BIGHORNS \citep{Sokolowski2015}, LEDA \citep{Price2018}, MIST \citep{Monsalve2024}, PRIZM \citep {Philip2019}, REACH \citep{deLeraAcedo2022}, and SARAS \citep{Singh2018}.  These experiments are all based on individual antennas paired with receivers designed to either have the smoothest possible spectral responses, or to yield properties that can be well modeled and fit during data analysis.  Additional efforts using interferometers have been undertaken utilizing lunar occultation to enable comparison between the Moon's (approximately) blackbody spectrum and the background sky spectrum \citep{McKinley2018, McKinley2020} or novel beamforming strategies \citep{DiLullo2020, DiLullo2021}.  

The Experiment to Detect the Global EoR Signature (EDGES) has been under iterative development and operation since 2006. Over this period, the experiment has evolved but remains close to the initial EDGES-1 concept \citep{Bowman2008, Rogers2008, Bowman2010}, which demonstrated the potential of a dipole-based antenna connected with minimal path length to a receiver utilizing internal noise comparison switching (e.g. Dicke switching) to yield a spectrally smooth instrument response with low systematics.  In addition to pathfinder science demonstrations, the early EDGES-1 systems characterized RFI \citep{Bowman2010rfi} at various sites in the United States and at Inyarrimanha Ilgari Bundara, the CSIRO Murchison Radio-astronomy Observatory in Western Australia (WA).  The WA Observatory is the primary site for EDGES for long-term deployments and is also the location of the Murchison Widefield Array \citep{Lonsdale2009} and the Square Kilometer Array-Low \citep{Garrett2010}.  

The design of EDGES instruments matured between 2010 and 2016 into EDGES-2, adding absolute calibration of the receiver in the laboratory based on scattering matrix and noise wave formalisms \citep{Rogers2012, Monsalve2016, Monsalve2017a}.  This was done to further reduce spectral structure in the calibrated output and enable fitting of foreground models with more physical realism during analysis.  EDGES-2 introduced a further simplified single-polarization ``blade'' dipole antenna that was deployed on successively larger ground planes to lower sensitivity to the soil beneath the antenna and reduce frequency dependence in the antenna power pattern \citep{Mahesh2021}.  It yielded useful astrophysical constraints on reionization and properties of early galaxies \citep{Monsalve2017b, Monsalve2018, Monsalve2019} and evidence for the first detection of the 21-cm signal \citep{Bowman2018}, although \cite{Singh2022}~argue that observations from SARAS-3 are inconsistent with that detection.  EDGES-2 also demonstrated passive characterization of the thermal electron temperature in Earth's ionosphere \citep{Rogers2015}, measured the average spectral index of the Galactic foreground \citep{Mozdzen2017, Mozdzen2019}, and constrained the amplitudes of diffuse radio recombination lines away from the Galactic plane \citep{Vydula2024}.  

The current EDGES-3 system implements a new antenna design with embedded electronics to remove the need for a balun between the antenna and receiver, similar in practice to the SARAS-3 antenna coupling \citep{Jishnu2021}.  EDGES-3 further enables in-situ absolute calibration by integrating the calibration standards used in the laboratory into the antenna electronics.   These hardware upgrades are coupled with new open-source analysis tools to provide robust, simultaneous parameter estimation for signal, foreground, and instrument models \citep{Murray2022}.  The EDGES-3 system has been tested in southeast Oregon in the United States, and was deployed to Devon Island in the Arctic for a one-month campaign in August 2022, and to Adak Island from December 2024 to April 2025.  These three deployments utilized temporary ground planes implemented as long meandering wires.  Since November of 2022 an EDGES-3 system has been operating at the primary EDGES site in WA, where it was deployed on a 50$\times$48.8~m (tip-to-tip) welded wire mesh ground plane, the largest used for EDGES to date.

In this paper we summarize the lessons learned from field experiences in WA pertaining to the challenges presented by this difficult measurement, and the stringent physical and instrumentation requirements that must be met in order to achieve the needed measurement precision (\S~\ref{sec:challenges}).  We then review the EDGES-3 instrument design and hardware status (\S~\ref{sec:edges3}), and conclude with results from laboratory testing and commissioning the EDGES-3 system in WA (\S~\ref{sec:commissioning}).  Further details on the instrument and deployments can be found in the Haystack EDGES Memo (HEM) series\footnote{https://www.haystack.mit.edu/haystack-memo-series/edges-memos/}, some of which are referenced in the following sections.


\section{Experimental Challenge} \label{sec:challenges}

The fundamental challenge for detecting the global 21-cm signal is to differentiate a weak ($\sim$~10~-~100~mK), broad ($\sim25$\% fractional bandwidth) spectral feature against a spectrally smooth foreground of 100~to 10,000~K and in the presence of other potentially spectrally structured contaminating influences.  Addressing this callenge necessitates an instrument with an extremely well-understood spectral response.  Many factors can contribute to spectral structure in a radio instrument, often at substantially higher levels than the one part in $10^{4}$ to $10^{6}$ required to separate the global 21-cm signal from foregrounds with an adequate signal-to-noise ratio (SNR) and high confidence.   

Typical approaches to end-to-end instrumental calibration for radio instruments operating at centimeter wavelengths and shorter are not easily applied at meter wavelengths for the global 21-cm measurement.  A blank reference field on the sky is not possible because the global 21-cm signal fills the entire sky.  Alternatively, an external calibrator made from radio-frequency absorber material (e.g. Eccosorb\reg) might be positioned in front of the antenna, presenting it with a blackbody spectrum at a known temperature. This method has been used successfully for absolute temperature measurements at shorter wavelengths with COBE FIRAS \citep{Fixsen1994} and ARCADE \citep{Fixsen2006, Fixsen2011}. However, this technique would require a calibration target size that scales up from 140 mm (i.e. 14 wavelengths) at 30 GHz (as in \citealt{Fixsen1994}) to 84 m (i.e. 14 wavelengths) at 50 MHz, and would still only provide a 300 K reference temperature, lower than the typical sky temperature by one to two orders of magnitude.  Connecting the antenna to a lamp filament to provide a 1600 K reference temperature was tested in HEM \#104 \citep{Memo104}, but for high-precision, the temperature of the calibrator must be regulated to closely match the sky noise temperature, which is unrealistic for the 100~to 10,000~K typical sky temperature seen below 200~MHz.  Further, over the wide bands needed for global 21-cm measurements, the beam pattern of an antenna or feed will vary with frequency, coupling angular structure in the Galactic foreground into spectral structure in the measurement.  External calibrators cannot address this source of spectral structure.  

In the face of such challenges, the strategy adopted by EDGES and most global 21-cm experiments is to break the problem into solvable parts.  Internal calibration sources can be used within a receiver to provide accurate calibration of the signal path after the antenna.  Implemented successfully, this approach shifts the remaining calibration challenge to the antenna and a localized electrical interface between the antenna and receiver.   Below we discuss considerations for the design and deployment of an antenna, with a summary of the various challenge mitigation strategies presented in Table \ref{tab:mitigation}.

\begin{table*}
\centering
\caption{EDGES-3 Mitigation of Observational Challenges} 
\label{tab:mitigation}
\begin{tabular}{l p{12cm}}
        \hline
	Challenge & EDGES-3 Mitigation Strategy \\
	\hline
    External Reflections & Locate in a flat, open site with a large ground plane.  Utilize an antenna design with a beam pattern that has low gain at the horizon. \\
    Internal Reflections & Use a precise calibration formalism in conjunction with multiple calibration sources and a VNA. \\
    Self-induced EMI & Employ internal shielded boxes for all electronics, use fiber optics where possible, and place ferrites on all current-carrying wires. \\
    Chromatic Ground Loss & The deployment of a large, conductive, welded wire-mesh ground plane. \\
    Chromaticity & Utilize a small antenna ($\lesssim$~1 wavelength) atop a large ground plane with serrated edges. \\
    Ionospheric Emission & Choice of a site with relatively minor ionospheric activity coupled with low-order ionospheric modeling. \\
    Human-generated RFI & Deploy the system to remote observation locations ($\gtrsim$~2000 km from the nearest FM transmitters). \\
    Solar Radio Emission & Restrict feature-detection analysis to deep night-time data, when the sun is $>$ 20\degree~below the horizon. \\
    Galactic Radio Recombination Lines & These can be ignored, as they are only a $\sim$~10 mK effect \citep{Vydula2024} \\
	\hline
\end{tabular} \\
\vspace{2mm}
This table provides a summary of the major challenges of detecting the global 21-cm absorption signal from cosmic dawn with a single instrument.  These challenges are examined in detail in \S~\ref{sec:challenges}.
\end{table*}


\subsection{Scattering and Reflections} \label{sec:scattering}

``Multipath'' reflections and scattering of incoming electromagnetic waves before entering the antenna yield frequency-dependent constructive and destructive interference in the spectrum.  These correlated signals typically manifest as sinusoidal ripples but higher-order structures may occur depending on the complexity of the environment.  Scattering can occur wherever there are changes in the impedance along the propagation path of the incoming waves, including at nearby structures and vegetation.  The period of the induced spectral ripple scales inversely to the relative propagation delay between the reflected or scattered signal path and the direct path.  Even relatively distant scatterers can lead to spectral ripples with periods similar to the scales of interest for global 21-cm measurements.   

An analytic expression to estimate the magnitude of the spectral ripples can be found in \cite{Rogers2022}.  In order to mitigate this effect, the area around the antenna must be kept clear of any large scattering objects.  Furthermore, since potentially problematic scattering objects will tend to be at low elevation, minimal antenna horizon gain is valuable for mitigating this source of systematic structure in the data.  Note that in the absence of a conductive ground plane, scattering from subsurface soil layers due to variations in electrical properties (e.g. buried boulders and bedrock) can also produce distortions in the measured spectrum. 

Similarly, reflections of electromagnetic waves within the physical structure and/or circuitry of the antenna and receiver can introduce interference structure into the spectrum.  Even compact dipole antennas can have significant internal delays that increase their spectral structure \citep{Mozdzen2016}.  Other reflections can occur at transitions between cables, adapters, and circuit components.

\subsection{Resistive Loss and Chromatic Ground Loss}
\label{sec:groundloss}

Frequency-dependent resistive losses can occur in the antenna, ground plane, transmission lines, and other components of the receiver since they are not perfect conductors.  In addition, any downward-facing component of the antenna beam is not sensitive to the astronomical signal and adds thermal power from the ground.  This effect is typically included in instrument response models as a resistive loss in the signal path, referred to as chromatic ground loss because it varies with frequency.  Such structure generally changes with time due to varying temperature and soil moisture content.  While in principle it is possible to model the ground loss with comprehensive measurements of subsurface conditions, doing so with high precision is a challenging prospect in most circumstances \citep{Spinelli2022, Monsalve2024}.  A highly reflective metallic ground plane under the antenna can electromagnetically isolate the antenna from the ground, with the isolation increasing with ground plan size (see \S\S~\ref{sec:chromaticity}, \ref{sec:welded},~and~\ref{sec:WA_ground}, as well as HEM \#370, \citealt{Memo370}).

\subsection{Chromatic Antennas and Sky Structure}
\label{sec:chromaticity}

Any broadband antenna will have frequency-dependent variations in its power pattern, known as ``chromaticity''.  When this chromatic beam pattern is convolved with the highly spacially structured galactic synchrotron emission it can produce frequency structure in the measured spectrum \citep[e.g.][]{Mozdzen2016, Sims2023}.  Minimizing this chromaticity in the band of interest and ensuring the antenna gain response changes in frequency and space as smoothly as possible allows for simpler modeling and less sensitivity to uncertainties in the antenna and sky models.  
 
The ability to model and remove spectral distortions caused by beam chromaticity is limited by imperfect knowledge of both sky and beam.  Even if the chromatic beam is known perfectly, convolution of that beam with an imperfectly known sky may still result in residual spectral distortions that cannot be accurately modeled.  Presently, no method exists to empirically measure a two-dimensional antenna beam pattern to sufficient accuracy to replace antenna models in analysis and calibration.  Considerable effort is underway to develop parameterized antenna and sky models able to account for unknowns \citep{Anstey2021, Anstey2022, Scheutwinkel2022, Pagano2023, Cumner2024}.  An achromatic beam, on the other hand, would eliminate spectral distortions from this beam/sky interaction regardless of imperfect sky knowledge.  Therefore, this source of systematics can be mitigated by designing instrumentation that minimizes absolute beam chromaticity. 

Under ideal conditions, electrically small ($\lesssim1$~wavelength) antennas have little intrinsic chromaticity, or only slowly varying frequency dependence, in their beam patterns and impedance, making them popular choices for experiments.  Larger antennas tend to have higher beam chromaticity in the frequency range of interest.  There is no sensitivity penalty for using small, low-gain antennas for global 21-cm measurements since the all-sky signal fills the response of any antenna. However, some analysis methods can capitalize on the smaller beam patterns of larger antennas to acquire more independent samples of the sky and better separate foregrounds from the expected 21-cm signal \citep{Liu2013, Anstey2023}. 

If a conductive ground plane is employed to isolate the antenna from non-uniform subsurface conditions, the ground plane itself along with the boundary of that ground plane with the surrounding soil become part of the antenna structure, leading the combined structure to no longer be electrically small, regardless of antenna design.  This added conducting plane will change the chromaticity, generally increasing variations of the beam with frequency, but potentially reducing the overall amplitude of those variations.  This effect can also be understood in terms of radio wave reflections from the discontinuity at the edge of the ground plane, which depend on the soil properties outside the ground plane.  

Beam chromaticity issues are addressed in HEM \#370 \citep{Memo370}, which shows that, in the EDGES case, the effects of beam chromaticity due to a ground plane are low for very small ground planes, reach a maximum for a ground plane 1.1 meters in diameter, and drop quickly for larger sizes.  The associated amplitudes of this effect vary depending on the electrical properties of the soil.  The amplitude of the ground plane-induced chromaticity is minimized by low antenna horizon gain.  It can be expected to decrease by a factor of $\sim$~4 for each doubling of the ground plane diameter.  For a rigorous analysis of the different ground planes used by EDGES, see \cite{Mahesh2021}.  


\section{Design of EDGES-3} \label{sec:edges3}

\begin{figure}
    \centering 
    \includegraphics[width=1\linewidth]{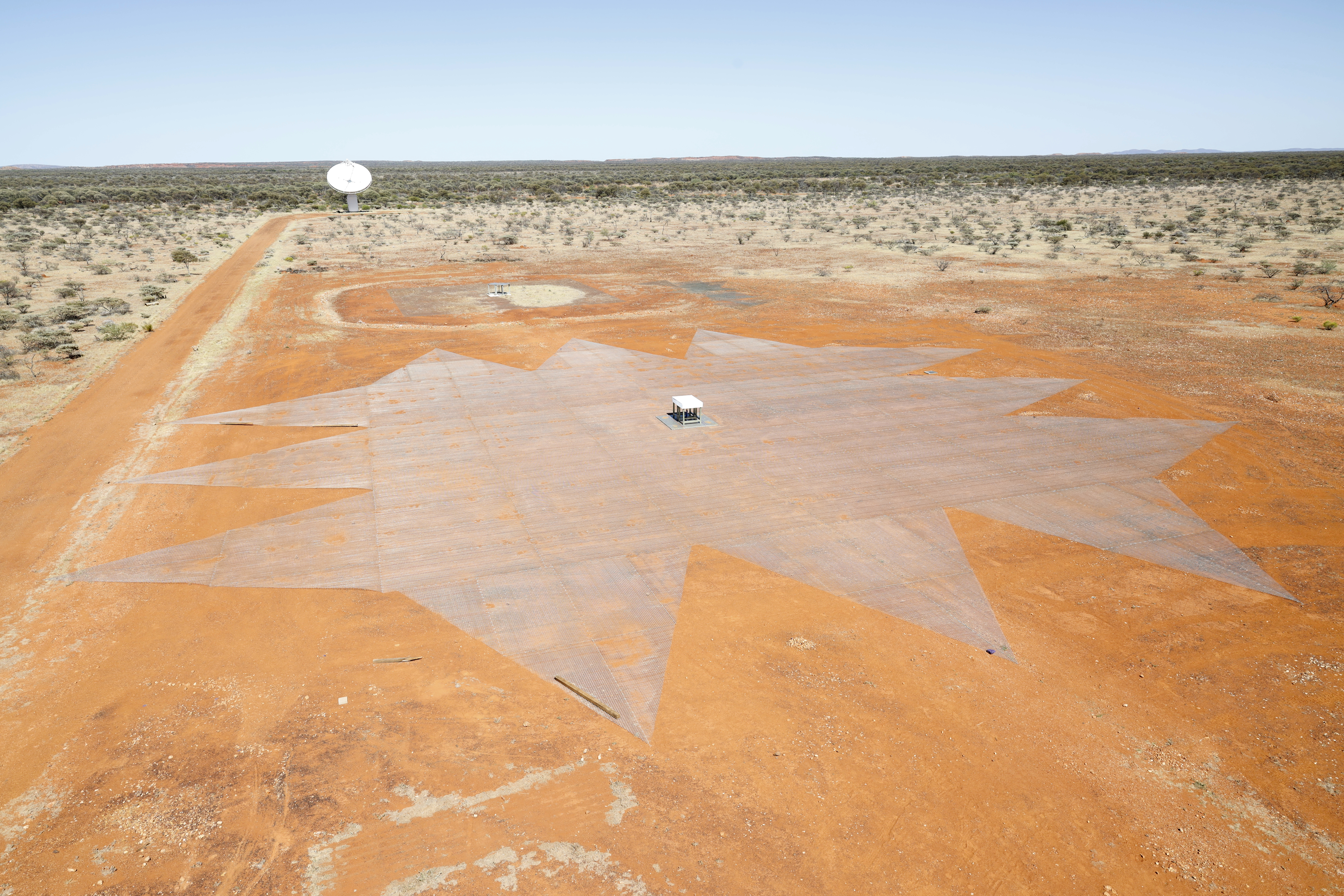} 
    \caption{Photograph of the EDGES-3 ground plane right after deployment (foreground), with the box-blade antenna at the center.  The smaller structure behind is the EDGES-2 low-2 ground plane and antenna used in \citet{Bowman2018}, and an ASKAP 12~m dish is visible in the distance.  After the photo was taken, the low-2 ground plane was enlarged to the same size as the EDGES-3 ground plane with the tips of each ground plane's triangular extensions nearly touching.}
    \label{fig:EDGES3-ground-plane}
\end{figure}

The EDGES-3 instrument was installed at Inyarrimanha Ilgari Bundara, CSIRO's Murchison Radio-astronomy Observatory\footnote{https://www.csiro.au/en/about/facilities-collections/atnf/mro} in Western Australia (hereafter referred to as the WA Observatory), in November of 2022.  The instrument is located near earlier EDGES instruments at -26.714853$^\circ$~latitude, 116.603968$^\circ$~longitude.  It consists of a large, conductive ground plane with a horizontally oriented box-blade dipole antenna centered on it (Figure~\ref{fig:EDGES3-ground-plane}). The dipole antenna consists of two hermetically sealed boxes that form the two arms of the dipole.  One of the boxes is empty while the other contains all analog and digital electronics. This design addresses the challenges identified in Section~\ref{sec:challenges} and improves on EDGES-2 on multiple fronts.  Like earlier EDGES systems, EDGES-3 covers roughly a 2:1 frequency range, as a reasonable trade-off between bandwidth and calibration accuracy.  By incorporating all electronics within the antenna box, the EDGES-3 design eliminates the need for the balun transformer used in previous systems, thus removing balun loss and reducing the delay in the antenna S11\footnote{For a one-port device, such as an antenna, the S11 and reflection coefficient are equivalent and in this paper both terms are used interchangeably.}, which in turn lowers the impact of uncertainties in antenna and receiver reflection coefficient measurements \citep{Monsalve2017a}.  The integrated electronics also include absolute calibration standards and a radio-frequency (RF) switching network to enable in-situ calibration, rather than pre- and post-observation calibration in the laboratory, as was done for EDGES-2.  

For EDGES dipole-like antennas on large ($>$~10~m) ground planes, chromatic beam effects above $\sim100$~MHz are dominated by the properties of the antenna and can be reduced by using smaller antennas closer to the ground.  Below $\sim100$~MHz, the beam chromaticity is dominated by edge effects from the ground plane and is reduced by increasing the size of the ground plane.  These considerations motivated a decrease in the EDGES-3 antenna size to approximately 75\% of the EDGES-2 low-band antenna, as well as an increase in the size of the ground plane to 50$\times$48.8~m, compared to 30$\times$30~m in the final version of EDGES-2.  These changes are predicted to lower the overall beam chromaticity and sensitivity to soil properties.  Simulations show that the EDGES-3 antenna and ground plane will reduce the average chromatic beam effect below 90~MHz across all local sidereal times (LSTs) nearly by a factor of two compared to EDGES-2, using the metric of the residual root mean square (RMS) after fitting a 5-term foreground model.  This result is within 33\% of the limit of an ideal, infinite ground plane.  For an in-depth analysis of the chromatic effects of different ground plane sizes, see HEM \#370~\citep{Memo370}.

\subsection{Welded Mesh Ground Plane} \label{sec:welded}

At the experiment's long-term site at the WA Observatory, EDGES-3 uses a 50$\times$48.8~m ground plane.  The ground plane is formed from galvanized steel square-grid mesh sheets that are 5$\times$2.4~m~in size with 50~mm grid spacing and 4~mm wire diameter.  The mesh sheets are welded together approximately every 200~mm along their perimeters.  Within each mesh sheet, the intersections of all wires are welded during fabrication by the vendor prior to galvanization.  The ground plane yields negligible transmission of radiation below 100~MHz.  This transmission is generally known as shielding loss, and electromagnetic simulations with the FEKO\footnote{\url{https://altair.com/feko}} software show that it is of order 0.016\%~for the EDGES-3 system, as reported in HEM \#316 \citep{Memo316}.  The welded, wire-mesh ground plane also offers structural rigidity for spanning small-scale ground irregularities.  This rigidity allows the mesh to rest directly on the ground, thus minimizing reflections from the mesh/soil boundary compared to elevated ground planes.  The shape of the ground plane is based on an inner 30$\times$28.8~m region formed with 72~mesh sheets surrounded by an outer boundary of three triangular extensions on each side, yielding a serrated (or perforated) overall shape.  Each triangle extension is 10$\times$9.6~m and created from four mesh sheets, two of which have been cut diagonally to yield four angled sections.  The triangular extensions are designed to minimize the effects of edge reflections \citep{Mahesh2021}.  

\begin{figure}
    \centering 
    \includegraphics[width=1\linewidth]{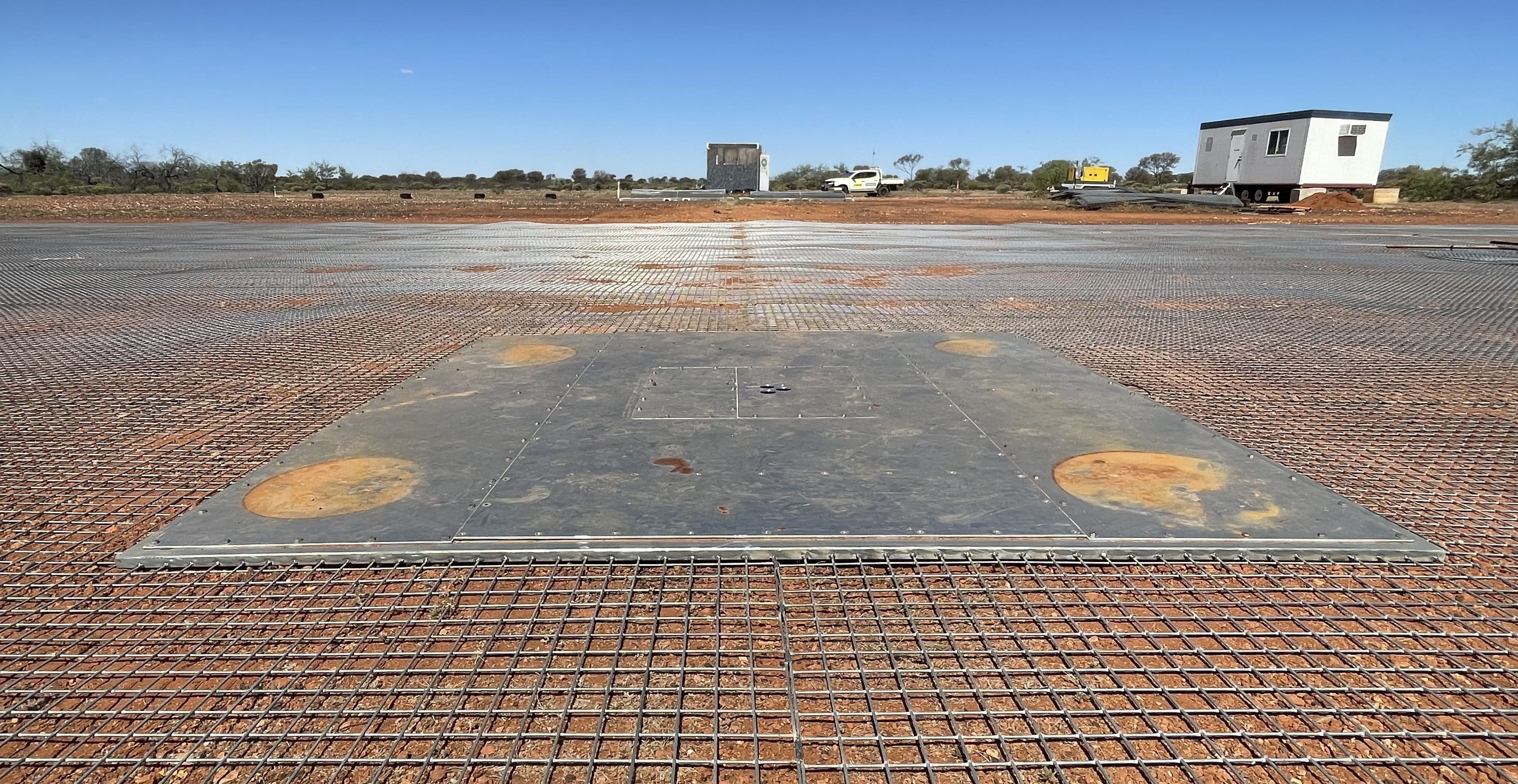} \\
    \includegraphics[width=1\linewidth]{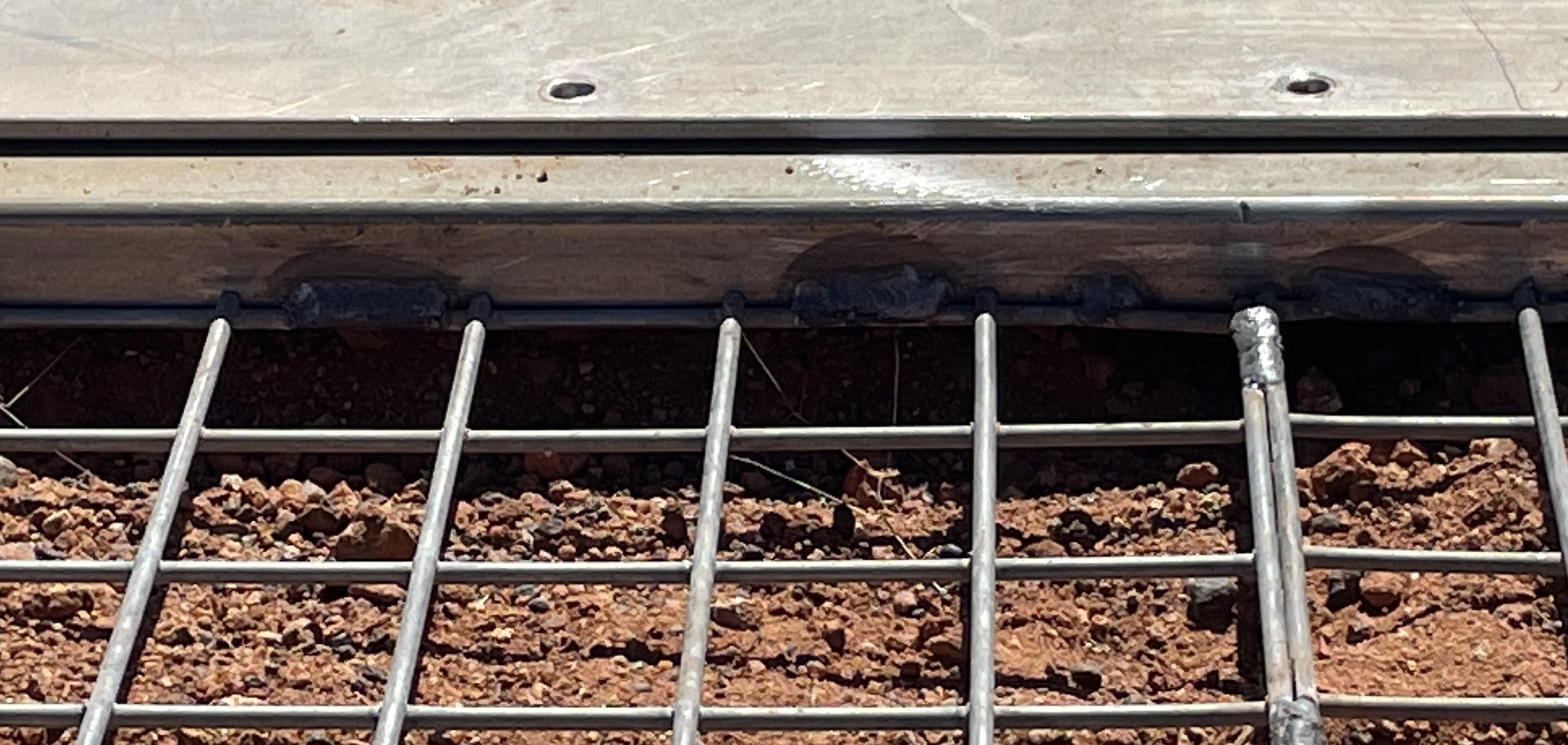}
    \caption{\textit{Top:}~Photograph of the EDGES-3 base plate before installation of antenna.  The base structure is welded to the neighboring mesh sheets every 100~mm on sides where the closest wire runs parallel to the structure (closeup shown in \textit{bottom} panel). Where the mesh wires intersect the base structure perpendicularly, every individual wire is welded to the base structure, working out to $\sim$~50-mm intervals.  Three aluminum plates are bolted to the top of the base structure with bolts spaced $\sim$~150~mm in the photo, but later additional bolts were added to lower the spacing between bolts to 75~mm, as discussed in Section~\ref{sec:75_res}.}
    \label{fig:EDGES3-base-structure}
\end{figure}

The central 2.4$\times$2.4~m region of the ground plane directly under the box-blade antenna is a base support structure for the antenna (top panel of Figure~\ref{fig:EDGES3-base-structure}).  This base is constructed from three 5~mm thick solid aluminum plates that were originally bolted every 150~mm onto a frame made from 75$\times$25~mm rectangular steel channel\footnote{After commissioning, a consistent 75 MHz feature was present in the data, the cause of which was tracked down to a slot resonance created by this 150~mm bolt spacing.  Extra bolts were added to push this resonance out of the band of interest.  For more details see \S~\ref{sec:75_res}.}.  The channel rests on the ground so that the total height of the central base structure is 30~mm.  A 707$\times$707~mm opening was cut from the aluminum plate at the center of the structure, providing access to underground PVC conduits buried $\sim$~0.5~meters below the surface that house electrical, fiber network, and forced air supplies.  Two small aluminum panels (one 707$\times$406~mm and the other 710$\times$301~mm) cover the access area when the instrument is operating.  They are bolted to a 25~mm lip extending into the cutaway on all sides under the surrounding aluminum plate.  The lip is welded to the bottom of the aluminum plate.  

The mesh sheets adjacent to the base structure were cut to match the outline of the base structure and welded to the exposed steel channel at ground level.  This design results in the top of the base structure protruding approximately 25~mm above the rest of the ground plane (bottom panel of Figure~\ref{fig:EDGES3-base-structure}), which was found to be a good mechanical design without added loss or beam chromaticity, as described in HEM \#328 \citep{Memo328}.  Details of the on-site assembly and construction of the ground plane are in HEM~\#406 \citep{Memo406}.  The EDGES-3 base structure is nearly 10\% larger than the 2.2$\times$2.2~m base structures used in EDGES-2 ground planes.  The welded mesh connections to the EDGES-3-base structure are designed to provide better electrical connection than EDGES-2, where mesh sheets were clamped between the channel frame and aluminum panels.  Together, these changes are intended to decrease possible resonances that could be introduced at the boundary between the base plate and the larger ground plane \citep{Bradley2019}.

Care is needed in construction of welded mesh ground planes, particularly at junctions between adjacent sheets.  It is possible to inadvertently create instances of closely parallel wires that have no electrical connection for a significant distance but substantial capacitance between them.  This can lead to resonances that affect the system response.  Such adjacent sheets must be electrically connected, without lengthwise gaps longer than a small fraction of the wavelength of interest to avoid these slot resonances contaminating the observational band.  This effect is further examined in HEM \#209 \citep{Memo209}. Another issue that arises in the field is buckling of the mesh due to dimensional variations and thermally induced stresses.  If these deviations from a planar surface cannot be eliminated, they must be mapped, and their effects incorporated into data analyses and interpretation.  Generally, vertical imperfections of a couple of centimeters or less are not a significant concern (see HEM \#383, \citealt{Memo383}).  These possible issues with the ground plane are further addressed in \S~\ref{sec:WA_ground}.

\subsection{Box-Blade Dipole Antenna}

\begin{figure*}
    \centering
    \includegraphics[width=0.9\linewidth]{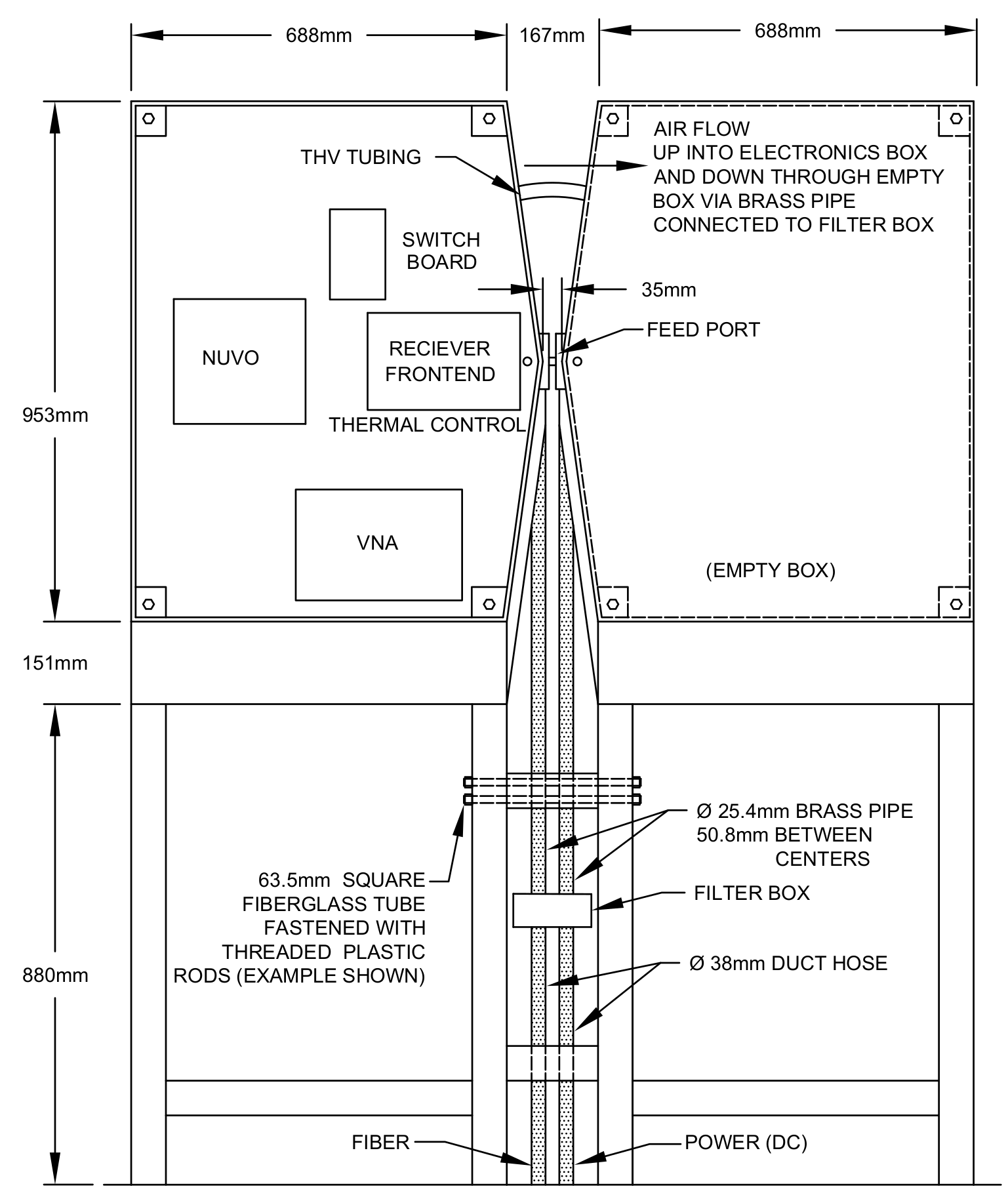}
    \caption{A fold-open schematic of the EDGES-3 antenna, with relevant dimensions.}
    \label{fig:EDGES3_schematic}
\end{figure*}

Sky radiation is collected by a wideband, dipole-like, box-blade antenna mounted horizontally above the ground plane. The dipole antenna consists of two hermetically-sealed rectangular boxes tapered at the feed.  A schematic of the antenna structure with the relevant labels and dimensions is given in Figure~\ref{fig:EDGES3_schematic}. Flat aluminum panels were cut for each side of the antenna boxes and mounted on an internal aluminum L-channel framework.  The outer antenna boxes are designed to provide $\sim$~100~dB isolation between interior and exterior (see HEM \#300, \citealt{Memo300}, and HEM \#425, \citealt{Memo425}).  Additionally, the analog and digital electronics sit within their own, smaller shielded boxes inside of the antenna boxes, leading to a total isolation of $\sim$~180~dB (HEM \#301~\citealt{Memo301}).  All exterior panels are bolted every 25~mm to reach this isolation (see \S~\ref{sec:shielding}).  The antenna boxes are supported on fiberglass legs, similar to earlier EDGES-2 systems.  A foam insulating cover was installed over the antenna boxes to reduce the thermal load during the hot daytime temperatures present in WA.

\subsubsection{Tuning Structure}

The antenna boxes are connected on the undersides with a tuning structure consisting of a transmission line of parallel hollow copper pipes (one from each antenna box, with diameters of 25.4 mm and whose centers are separated by 50.8 mm), shorted at the filter box, which is 340~mm below the undersides of the antenna boxes (about 40\% of the way to the ground plane).  Each pipe is connected to its antenna box through a small hollow metal box (Hammond\reg~1590G, 100$\times$50$\times$25~mm) that extends 25~mm below the antenna box and 12.5~mm into the region between the antenna boxes.  At their bottoms, the copper tuning pipes are connected both physically and electrically with an aluminum box.  The dimensions, spacing, and length of the copper tubes, along with the exterior dimensions of the antenna, were determined through electromagnetic simulation to optimize the antenna impedance and beam chromaticity between roughly 55~and 110~MHz.  

Two vertical plastic air hoses pass from underground conduits through the holes in the ground plane base structure and connect to the bottoms of the hollow tuning pipes.  An additional horizontal plastic hose connects the two antenna boxes, enabling a loop for dehumidified forced air to circulate between the antenna boxes and through a 50~m underground conduit to stabilize the internal temperature of the boxes. A fiber optic cable exits the underside of the antenna box containing the electronics and runs down into the base structure where it is patched back to the supporting electronics hut 50~m away that provides power and network connection from CSIRO's facility infrastructure.  A two-wire insulated copper power line from buried conduit extends vertically through the ground plane to the antenna tuning structure.  It is split into individual wires before entering the tuning structure alongside the forced air hoses.  Each conductor has a ferrite core installed after the split to dampen RF resonances.  The location of the fiber optic cable and air hoses are not critical as they contain no electrically conductive material.  The copper power wire pair is centered under the antenna. The split wires are run symmetrically into the horizontal bar in the tuning structure and up into the active antenna box to power the electronics.  The vertical orientation and symmetry ensure minimal disruption to idealized antenna properties.  EDGES-3 operates off a 12V DC power supply isolated from earth ground.

\subsubsection{Feed Port}

\begin{figure}
    \centering 
    \scalebox{-1}[1]{\includegraphics[width=\linewidth]{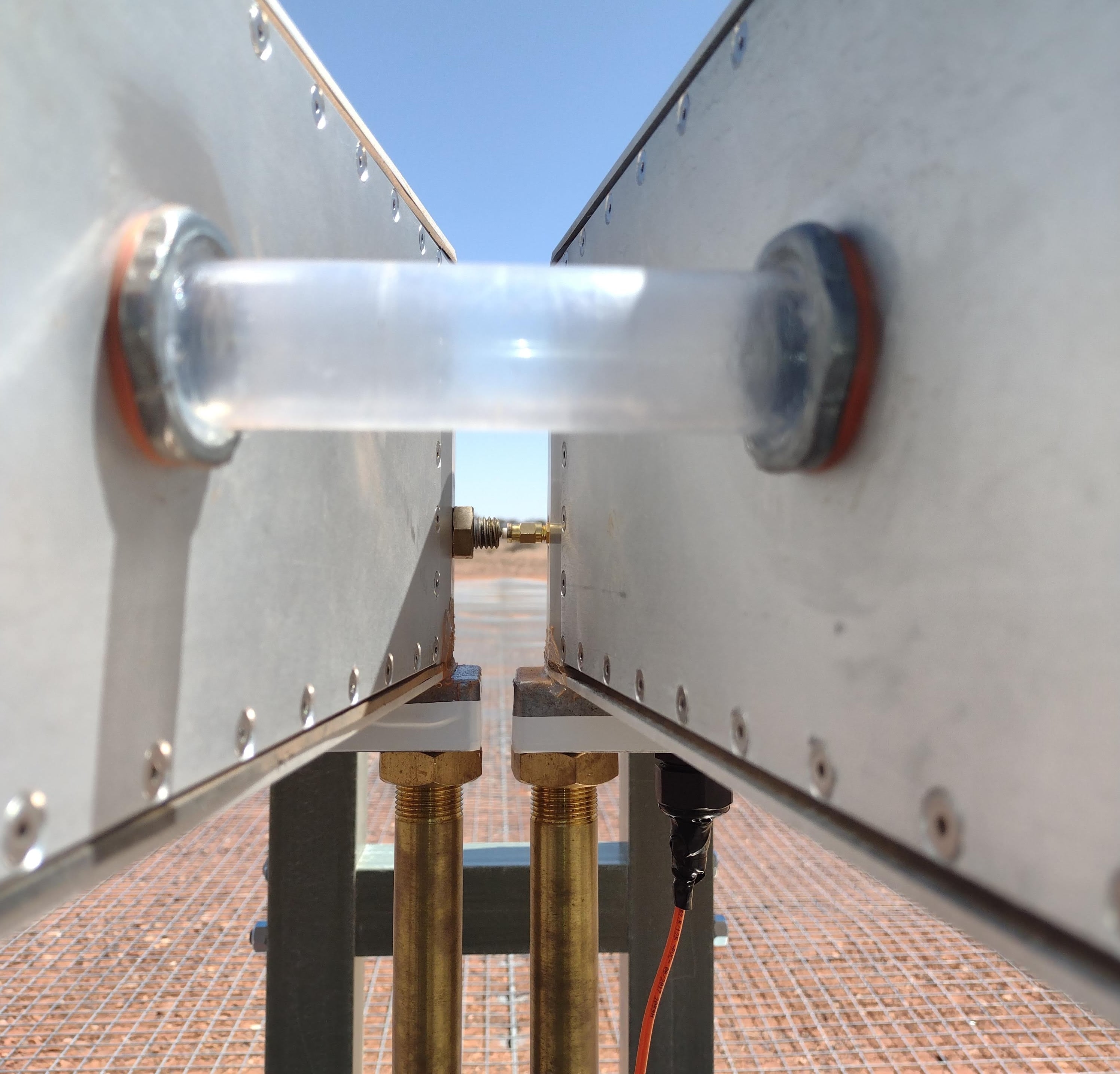}} \\
    \includegraphics[width=\linewidth]{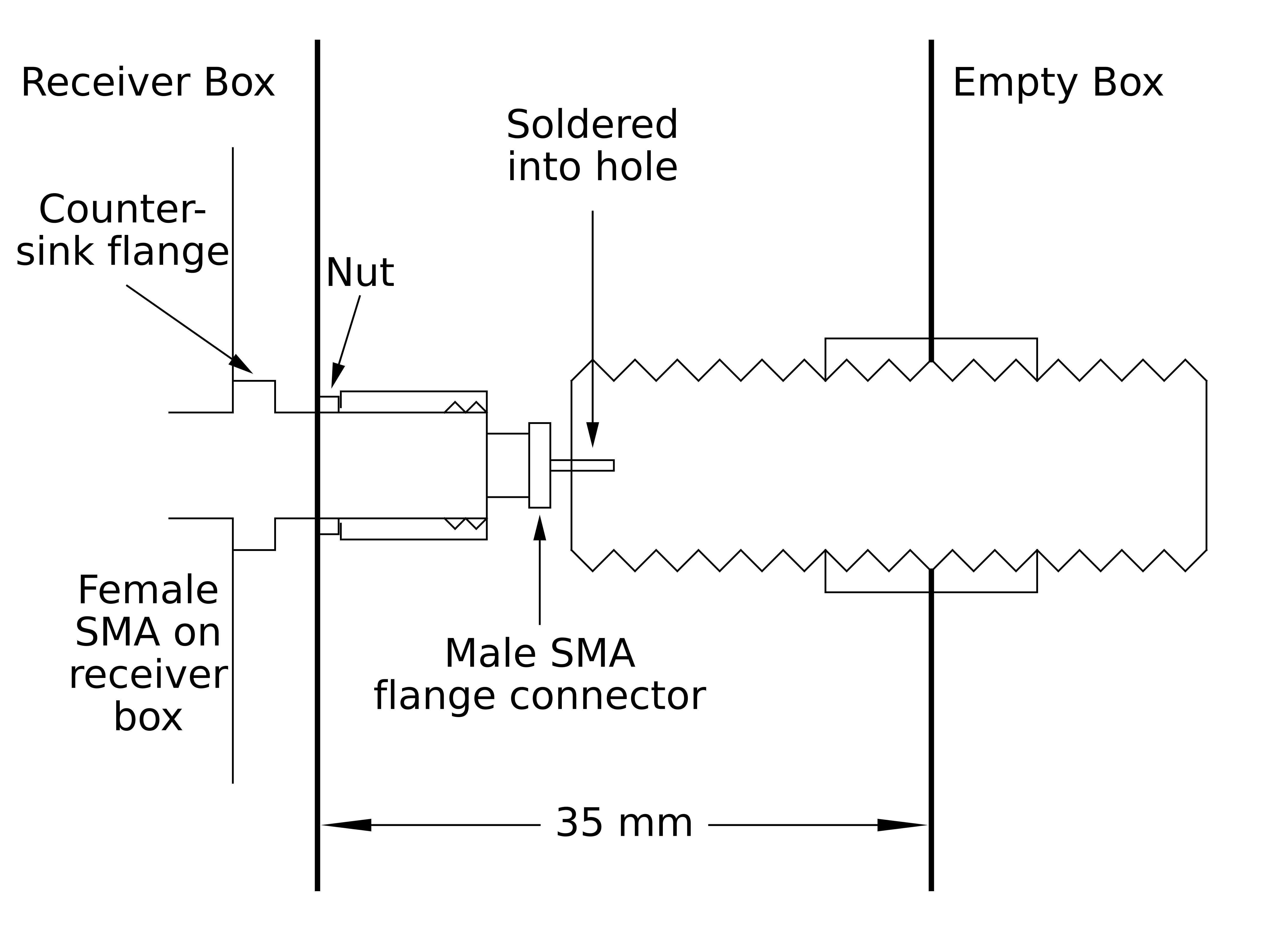}
    \caption{\textit{Top:}~Photograph of the radio-frequency feed port between the antenna boxes.  The hollow copper pipes in the tuning structure are visible under the feed port, extended down from the antenna boxes, and the plastic air-flow tube between the boxes can be seen in the foreground. \textit{Bottom:}~Schematic of the feed port between the two antenna boxes.  The coaxial connector is soldered into the hole in the center of the 12.7 mm diameter threaded brass rod, with nuts on each side of the empty box panel, tightened to provide an electrical connection to the empty box.}
    \label{fig:feed}
\end{figure}

The radio frequency feed port between the two antenna boxes (each antenna box forms one arm of the dipole) is comprised of SubMiniature version A (SMA) connectors.  The antenna box containing the electronics is connected to the ground of a female-female SMA bulkhead connector.  The inner conductor of a male SMA connector is soldered to the threaded extension.  The outer conductor of the SMA extension is removed.  To complete the feed, the male SMA port from the non-electronics box is connected to the female SMA port on the electronics antenna box.  The interior side of the female-female SMA bulkhead in the electronics antenna box is connected directly to the SMA male connector of the receiver input.  The top panel of Figure~\ref{fig:feed} is a photograph of the feed, which attaches via nuts to each box, and the bottom panel is a schematic of the feed port.

\subsubsection{Removal of the Balun}

Eliminating the balun, which in EDGES-2 introduced $\sim$~7~ns of propagation delay, reduces the total delay in the S11 phase of the antenna as seen by the receiver from $\sim$~24 ns to $\sim$~17 ns.  Simulations in HEM \#289 \citep{Memo289} show that, assuming a realistic 0.1~ns error in the measured antenna delay, removing the balun reduces residuals in a 5-term foreground model fit in simulated observations by a factor of $\sim$~2.  A similar improvement is seen if the measurement error is assumed to be for the reflection coefficient of the LNA input, rather than the antenna.  Removing the balun also eliminates its frequency-dependent~0.03~dB loss which was present in the EDGES-2 low-band.  Previously, this loss had to be modeled and removed from the observed spectra during analysis.

\subsection{Receiver}

\begin{figure*}
    \centering
    \includegraphics[width=0.9\linewidth]{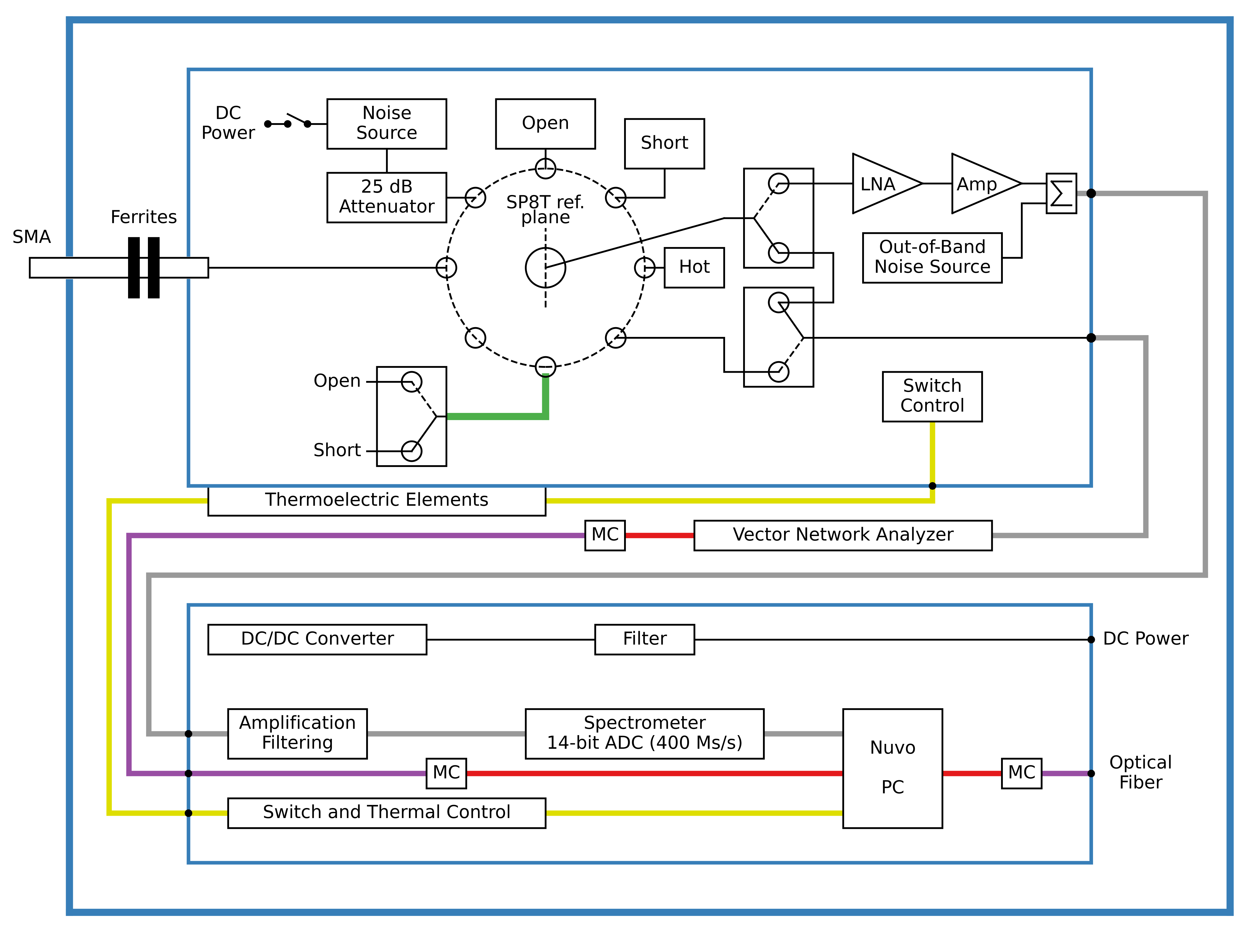}
    \caption{A diagram of the EDGES-3 signal chain.  Optical fiber is in purple, Ethernet cables are in red, coaxial cables are gray, MezIO control wires are in yellow, and copper wire is thin and black.  The long calibration cable is shown in green.  There are three media converters in the chain, each labeled MC.}
    \label{fig:EDGES3-signal-chain}
    \vspace{10mm}
\end{figure*}

\begin{figure}
    \centering
    \includegraphics[width=0.9\linewidth]{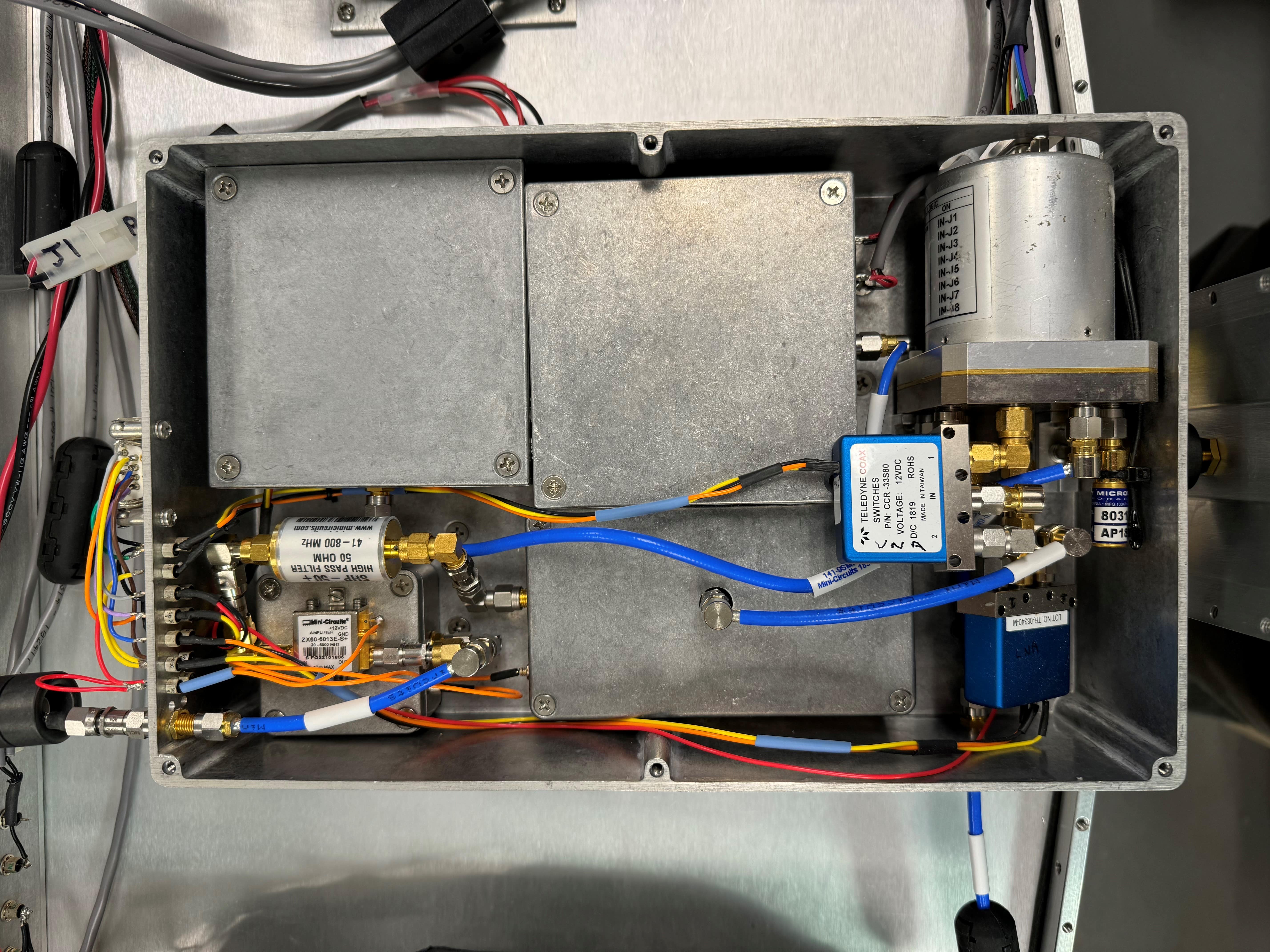}
    \caption{The EDGES-3 frontend box, which contains the long cable box, the box with the 50 ohm hot load, followed by the rotary switch along with the other components in the frontend box signal chain (details can be found in Section~\ref{sec:cal}).}
    \label{fig:fe_box}
\end{figure}

The EDGES-3 signal chain is closely derived from EDGES-2 and is shown in Figure~\ref{fig:EDGES3-signal-chain}.  A picture of the analog front-end is shown in Figure~\ref{fig:fe_box}.  It consists of a temperature-controlled, RF-shielded aluminum box that contains an RF~switching network with five calibration standards and internal comparison sources, a low-noise amplifier (LNA), a second-stage amplifier, and an out-of-band noise source.  A Keysight\reg~FieldFox N9923A Vector Network Analyzer (VNA) rests just outside the front-end shielded box, which has coaxial RF input ports for the antenna and the VNA, and an output port to a back-end signal conditioning chain and data acquisition system.  The box also contains 14 filtered pass-throughs for power and control lines.

The RF switching network is new for EDGES-3 (Figure~\ref{fig:EDGES3-signal-chain}) and formed from a Mini Circuits\reg~MSP8T-12D+ single pole, eight throw (SP8T) mechanical rotary switch and three Teledyn\reg~CCR-33S80 single pole, double throw (SP2T) mechanical latching relays.  Both switch models have low insertion loss ($<0.1$~dB) and high isolation ($>90$~dB). The network allows multiple measurement configurations to be realized: (1) connecting the LNA input to the antenna port or to any of the internal calibration and comparison loads, (2) disconnecting the LNA and instead connecting the VNA to the antenna port or any of the internal calibration loads, and (3) connecting the VNA to LNA input.  This switching network enables automated acquisition of antenna and internal comparison spectra in addition to reflection coefficients for stable science observing and periodic absolute calibration of the signal chain \citep{Rogers2012, Monsalve2017a}. 

\begin{figure}
    \centering
    \includegraphics[width=0.9\linewidth]{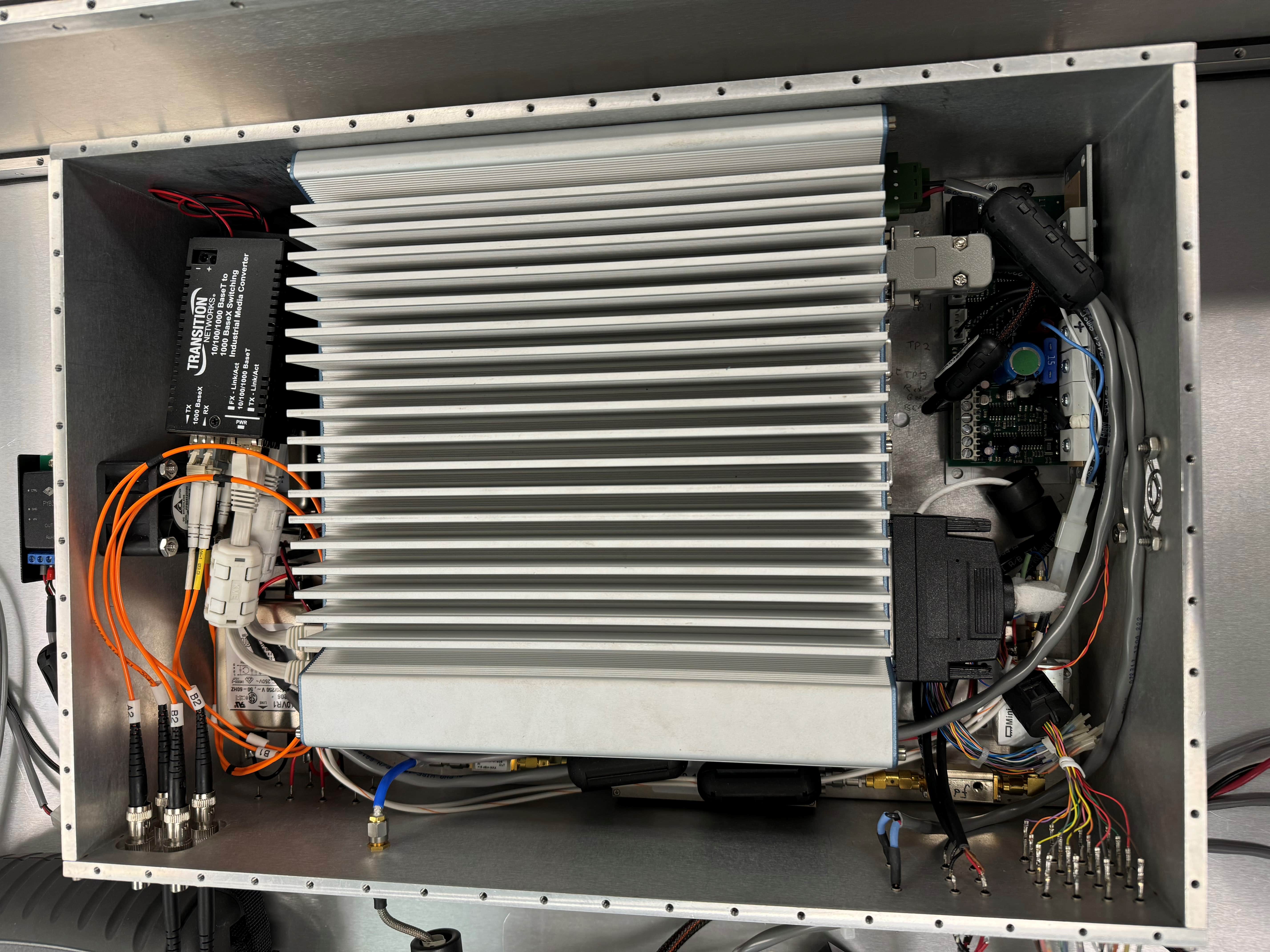}
    \caption{The shielded backend electronics box inside of EDGES-3, with the Novo~computer inside.  The various MezIO control wires can be seen exiting the box in the lower right, and the optical fiber leaves the box in the lower left, coming from the media converter visible in the upper left corner of the box.}
    \label{fig:inner_box}
\end{figure}

The digital back-end electronics consist of an RF-shielded box containing a Nuvo\reg~compact, fanless, industrial computer equipped with a 14-bit, 400~MHz Signatec\reg~PX14400 analog-to-digital converter (ADC) on an internal PCIe board for spectrometer data acquisition (Figure~\ref{fig:inner_box}).  The computer includes a MezIO\reg~digital I/O interface for control of the RF switches and thermal system, as well as an ethernet network connection to the VNA.  External connectivity into the back-end system computer for data acquisition and transfer is achieved through a network connection over optical fiber.  

During acquisition of spectra, the RF switching network in front of the LNA cycles successively between three states: the antenna, an internal matched comparison implemented as a 25~dB attenuator within the thermally controlled electronics box, and a ``hot'' reference enabled by powering on a Zener noise diode behind the attenuator, raising the effective noise temperature to $\sim$~1000~K.  The switch cycle time is configurable and typically set to dwell 13~seconds in each state, yielding a 33\% duty cycle for the antenna. The system is assumed to be stable on timescales within this cycle.  Raw voltage samples are acquired by the ADC continuously within each switch state, channelized with the \textsc{FASTSPEC}\footnote{http://github.com/edges-collab/fastspec} CPU-based polyphase filter bank (PFB) code, and accumulated to the desired integration.  The three spectra from each switch cycle are written to disk in a custom text file using Uuencoding.  The spectral resolution and PFB window functions are adjustable and typically configured to 6.1~kHz channels formed using a 5-tap Blackman-Harris window function in the PFB.  The antenna reflection coefficient (i.e. S11) is nominally measured by the VNA every 36~hours.  Thermal controller and temperature sensor readouts are recorded to disk every five minutes.  With these settings, the system generates $\sim$~1.5~GB of recorded data per day.

\subsubsection{RF Shielding}
\label{sec:shielding}

Electromagnetic leakage between different components of the instrument can lead to spectral ripples and more complicated structures.  Examples include leakage of an amplified received signal back into the antenna, between amplifier stages of the receiver, or within the digitizer, including feedback or correlations between time samples of the digitized signal.  Since all of the electronics of EDGES-3 are within the antenna, sufficient internal shielding is integral to preserving a smooth spectral response.  

A simple estimate of the necessary shielding that agrees well with more detailed calculations starts with a requirement to limit any leakage below the 21-cm signal level.  Using a conservative $T_{21}=10$~mK antenna temperature and $B=6$~kHz spectral channels, this limit is equivalent to a received channel power of $P_{21}=k_B T_{21} B = -180$~dBm.  The largest amplified RF signals inside the antenna box have RMS amplitudes of $V_{in}\sim1$~V, which in a 50~Ohm system yield an average amplified power of $P_{in}=V_{in}^2/R\sim13$~dBm.  If all such power is in a single spectral channel, a total isolation of $L_{total}=P_{in}/P_{21}=193$~dB is needed.  

\begin{figure}
    \centering
    \includegraphics[width=\linewidth]{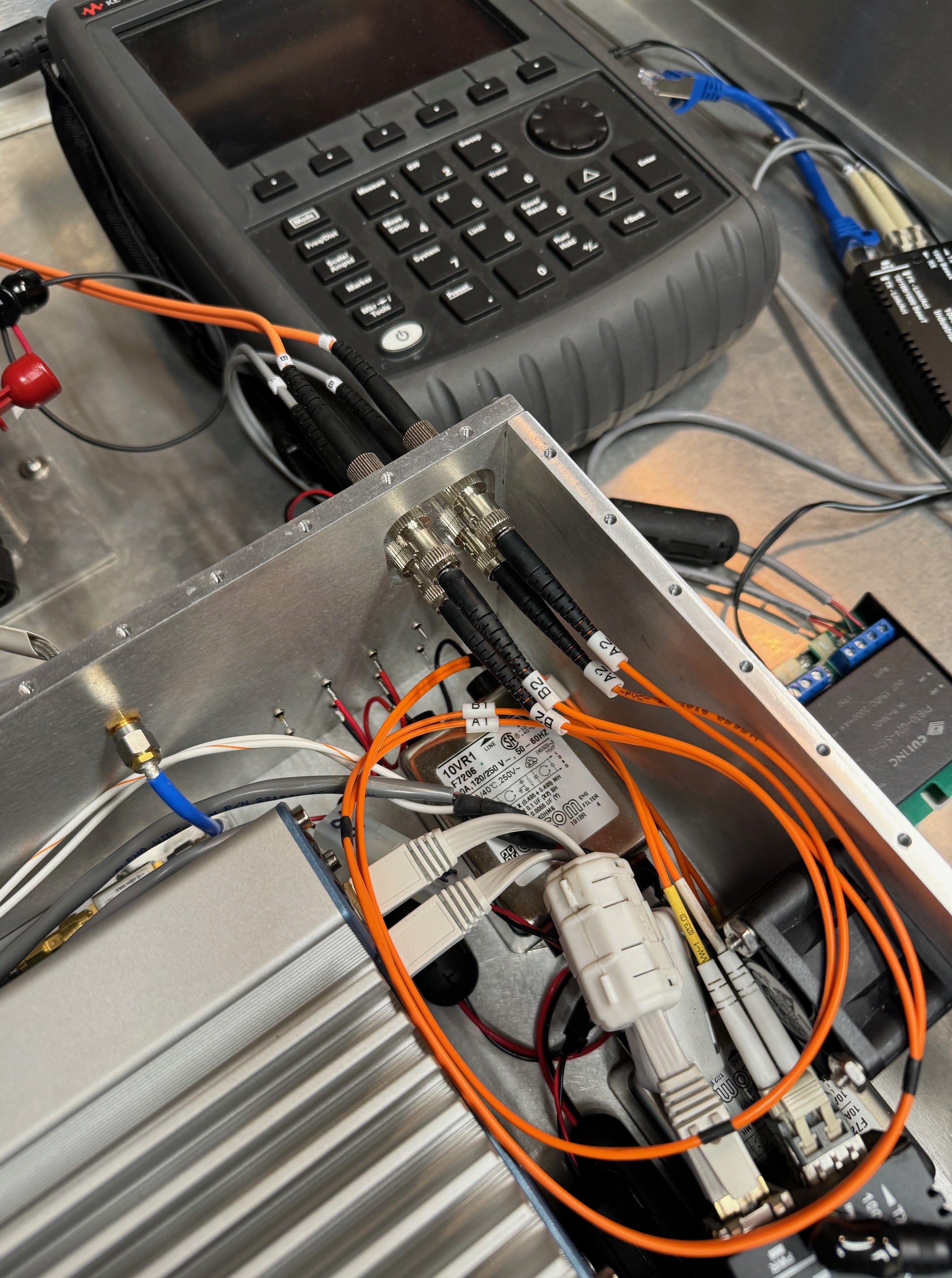}
    \caption{A closeup of the optical fiber leaving the shielded backend box.  The VNA and a media converter can be seen in the background.}
    \label{fig:VNA_conn}
\end{figure}

A typical coaxial cable provides about 100~dB of isolation.  Assuming internal shielding around the final stage of amplification provides similar levels of isolation, then the outer antenna box must provide the remaining $L_{box}\approx100$~dB.  For a metal box designed as a Faraday cage, slot resonances must be carefully avoided to ensure this level of isolation, usually achieved through any combination of welding seams, spacing bolts close together, and/or using RF gaskets to ensure high conductance across seams.  For the EDGES-3 antenna boxes, we found through laboratory testing that spacing bolts 25~mm (about 0.01 wavelength at the highest target frequencies), in addition to using a conductive nickel-graphite-silicone gasket at the seams, met the 100 dB additional isolation requirement.  To achieve the necessary additional internal shielding around electronics, the industrial computer (including its ADC) is placed within an inner aluminum box with similar 25~mm bolt spacing and gasket.  All power lines to this inner box are double filtered with ferrites and ethernet connections use optical fiber.  These connections are shown in the pictures of Figures \ref{fig:inner_box}~and~\ref{fig:VNA_conn}.  Further details and additional calculations for the shielding used in EDGES-3 can be found in HEMs~\#299, \#301, and~\#425 \citep{Memo299, Memo301, Memo425}. The Deuterium Array achieved 192~dB of isolation using similar enclosures \citep{Rogers2007}. 

\begin{figure}
    \centering
    \includegraphics[width=\linewidth]{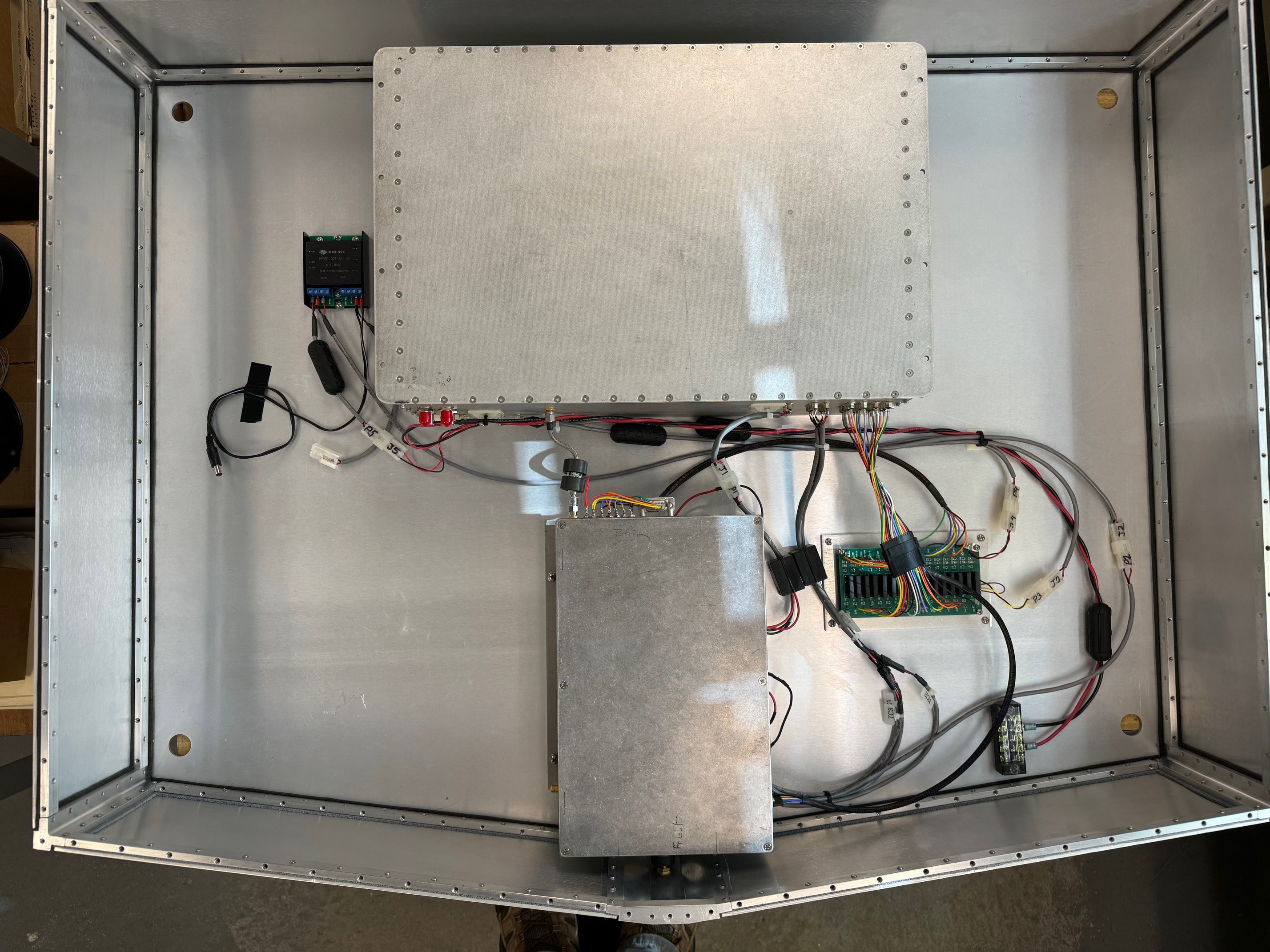}
    \caption{Interior of the EDGES-3 system, without the VNA and media converter, which can be seen in Figure \ref{fig:VNA_conn}.  The smaller internal box is the shielded front end (Figure~\ref{fig:fe_box}), and the larger internal box is the shielded back end, including the Nuvo~computer (Figure~\ref{fig:inner_box}).}
    \label{fig:EDGES3-inside}
\end{figure}

\subsubsection{In-Situ Calibration System}
\label{sec:cal}

\begin{figure}
    \centering
    \includegraphics[height=1\linewidth]{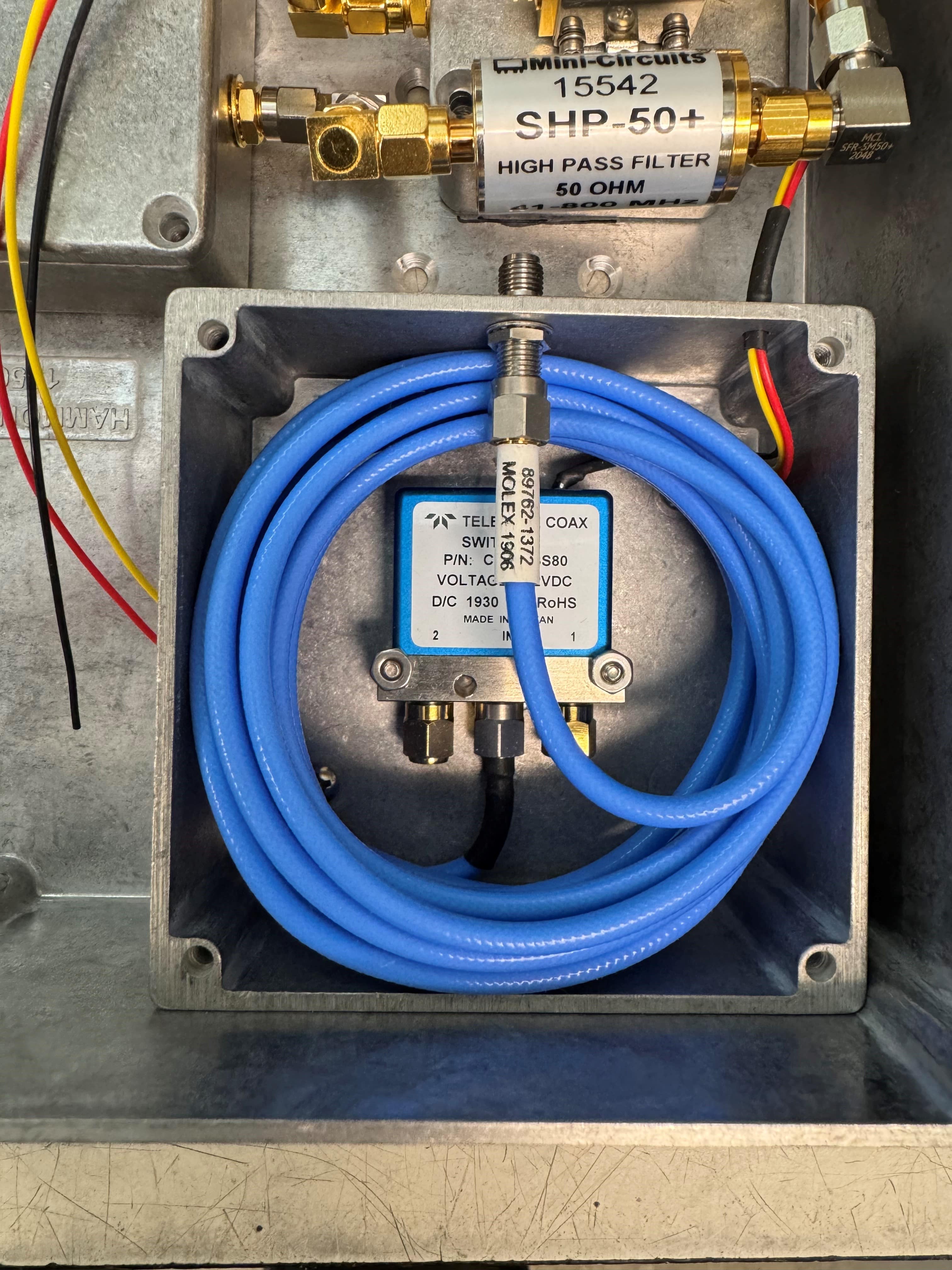}
    \caption{The EDGES-3 long calibration cable, a Molex\reg~89762-1372, coiled in its shielded box, shown during construction without the cable connection to the rotary switch.}
    \label{fig:EDGES3-cable}
\end{figure}

The manual absolute calibration process used for EDGES-2 in the laboratory was lengthy to perform and limited by repeatability errors due to the numerous connect/disconnect cycles needed to attach different calibration standards and devices to the receiver.  The EDGES-3 receiver replaces the manual process with an automated in-situ process enabled by the internal RF~switching network. Calibration spectra and reflection coefficient measurements are used to solve for the absolute receiver gain and noise temperature.  The absolute gain and noise temperature account for impedance mismatches between the antenna and receiver, as well as for correlated and uncorrelated LNA noise waves.  Three of the five internal sources are industry standards (Maury\reg~Microwave 8050 with offset delays of 33~ps) providing nearly ideal open, short, and matched 50~Ohm load references used to calibrate the VNA. The other two internal sources are: (1) a matched load and (2) a long cable.  The 50~Ohm matched load is contained in an insulated metal box and connected through a short coaxial cable to the RF switch.  It is used at two physical temperatures: the thermally-controlled front-end temperature (usually set to 25 or 30$^\circ$~C) and heated to $\sim100^\circ$~C to provide a ``hot load''.  The long cable is a 3.05~m Molex\reg~89762-1372, coiled in a metal box and connected at its far end to a third Teledyne\reg SP2T latching relay with open and short terminations on its ports (Figure~\ref{fig:EDGES3-cable}).  

To characterize internal signal path reflections, EDGES measures the reflection coefficient of the LNA input and antenna directly with the VNA.  In order to solve for all of the frequency-dependent quantities in the calibration equations, the 50~Ohm ambient/hot load provides the primary temperature references (both the ambient and hot states are needed), while noise-wave parameters are estimated from frequency-dependent measurements of the open/shorted cable (again both terminations are needed). For each of the four calibrator states, it is necessary to obtain: the uncalibrated temperature spectra, the physical or noise temperatures, and the reflection coefficients.  These twelve receiver calibration products are then used to determine the absolute receiver gain and noise temperature.  The algorithms used for this absolute calibration technique are summarized in HEM \#132 \citep{Memo132}, \cite{Rogers2012}, and \cite{Monsalve2017a}.  Similar techniques have been used in other microwave measurements \citep{Hu2004, Belostotski2011, deLeraAcedo2022}.   

In EDGES-3 the hot/ambient load and open/shorted cable are not placed at the receiver input reference plane as in EDGES-2.  The calibration standards installed on the internal SP8T switch have a shorter path to the LNA than the antenna connected to the receiver input (see again Figure~\ref{fig:EDGES3-signal-chain}).  This difference in path length needs to be accounted for in the calibration model as it adds extra delay and loss for the antenna.  We simplify the problem by defining the reference plane for calibration at the eight terminals of the SP8T switch where all of the calibration sources and standards are located.   The S-parameters of the paths from each of the SP8T terminals to the common port are identical, which we confirmed by VNA measurements on an isolated switch to within $\pm1$~ps delay and 0.001~dB amplitude.  We treat the extra path length from this reference plane to the receiver input as part of the antenna since the antenna S11 acquired by the system includes the path.  With this approach, we only need to additionally determine the S-parameters of this extra path in order to calculate the additional loss it adds to the antenna, which we include as a thermal noise contribution in the calibrated spectra with the other antenna losses.  The S-parameters of the extra path can be found by connecting open, short, and load standards to the receiver input and comparing their S11 acquired through the receiver with the S11 measurements of the internal open, short, and load standards.  More details on this calibration technique can be found in \S~\ref{sec:path_char}. 

All VNA measurements cycle through the open, short, and matched load calibration standards and the port under test sequentially and repeatedly to achieve long integration times for low noise.  This cycling minimizes potential errors due to thermal drift in the VNA, which is not in the temperature controlled front-end electronics box. The VNA is powered on at least 60~minutes prior to its use to reach its nominal operating temperature.  With the appropriate integration time, the VNA can obtain an S11 fractional error better than $10^{-4}$ (HEM \#363, \citealt{Memo363}).  The accuracy of the reflection coefficient measurements is improved by accounting for the actual resistance of the matched 50~Ohm load calibration standard, which differs by $\sim0.01$~Ohm in practice from the ideal 50~Ohm, and the added inductance in the standard due to skin effects in the few millimeters of transmission line between the reference plane and internal load \citep{Monsalve2016}.  In order to correct for losses in the hot load, full S-parameters of the short cable to its internal matched load are measured prior to installation into the EDGES-3 receiver and assumed to be stable thereafter. 

High calibration performance is realized by the system via the combination of an antenna reflection coefficient lower than $-14$~dB over the observing band and a custom high-electron-mobility-transistor-based LNA.  The LNA reflection coefficient is lower than $-30$~dB across the band, the correlated LNA noise is less than 30~K, and the uncorrelated LNA noise is less than 200~K (see~\S~\ref{sec:simulator}). The 14-bit digitizer, augmented by an out-of-band noise source to ensure all switch states sample nearly identical voltage ranges in the digitizer to mitigate any non-linearlity in its response, provides high dynamic range which allows the system to tolerate intermittent RFI without saturation.

\subsection{The Australian Long-term Operation Site}

EDGES-3 at CSIRO's WA Observatory addresses surface scattering (\S~\ref{sec:scattering}) by choosing a deployment location where the box-blade antenna is at least 50~m distant from any local scatterers.  The EDGES on-site support hut is approximately 2$\times$3$\times$3~m in dimension and is located 50~m east of the box-blade antenna.  The side of the hut facing the antenna has been covered in RF absorber tiles, although its reflective cross-section has not been evaluated.  The EDGES-2 low-2 instrument is centered 50~m west of EDGES-3.  During 2023, its ground plane was expanded to match the size of the EDGES-3 ground plane, which places the tips of the triangular extensions of each ground plane within $\sim1$~m of each other.  A 12~m diameter ASKAP \citep{Johnston2007} radio telescope dish is 300~m west of EDGES-3. There are no trees or bushes within 40~m in any direction and only sparse vegetation beyond.  In the far-field, the site has no hills or mountains on the horizon and only $\sim1^\circ$ variations above horizontal. For EDGES, modeling shows that a horizon that is 5\degree~above horizontal contributes $\sim$~300 mK and can be accounted for by the degrees of freedom in the foreground models used in parameter estimation, but a 10\degree~horizon contributes $\sim$~2.5 K and requires dedicated model parameters to facilitate removal (HEM \#94, \citealt{Memo94}).  Sub-surface scattering is addressed by including a large conductive ground plane as part of the antenna.


\section{Testing and Commissioning}
\label{sec:commissioning}

\subsection{In-Situ Calibration}
\label{sec:caltests}

EDGES-3 can perform in-situ calibration that tests the system by substituting the open and shorted cables as well as the hot and ambient loads as ``antennas'' in software.  Each of these should produce relatively flat spectra at the correct temperature, with low residuals.

\subsubsection{Initial Path Characterization}
\label{sec:path_char}

The automated calibration of EDGES-3 is based on defining the S11 reference plane at the circle of the 8 input ports of the SP8T switch (as previously described in \S~\ref{sec:cal} and shown in Figure~\ref{fig:EDGES3-signal-chain}). In this arrangement the short, open, and load calibration standards are located at the reference plane, but small additional corrections are required for the hot load loss and for the short (38 mm) input cable and its associated connectors, which sits behind the 8-position switch and the antenna feed.  This input cable loss can be absorbed into the antenna loss. A correction to the calibrated S11 for the LNA input is also needed to account for the different path the VNA measurement takes.  This correction is determined from a search of cable length, cable dielectric correction, and cable loss correction that minimizes the residuals to the calibration results for the open and/or shorted cables.  The cable properties are given in Table~\ref{tab:cable-properties}, with more information on this technique available in HEM \#303 \citep{Memo303}.

\begin{table}
\caption{Cable Properties} 
\centering
\begin{tabular}{ll}
\hline
Property & Value \\ 
\hline
Length & 38 mm \\
Inner conductor diameter & 0.9195 mm \\
Inner conductivity & $5.96 \times 10^7$ S/m\\
Outer conductivity & $4.768 \times 10^7$ S/m \\
Dielectric diameter & 2.9845 mm\\
Dielectric constant & 2.04 \\
Loss tangent & $2 \times 10^{-4}$ \\
\hline 
\end{tabular} 
\label{tab:cable-properties} 
\end{table}

\subsection{Laboratory Testing With Antenna Simulator} \label{sec:simulator}

\begin{figure}
    \centering
    \includegraphics[width=\linewidth]{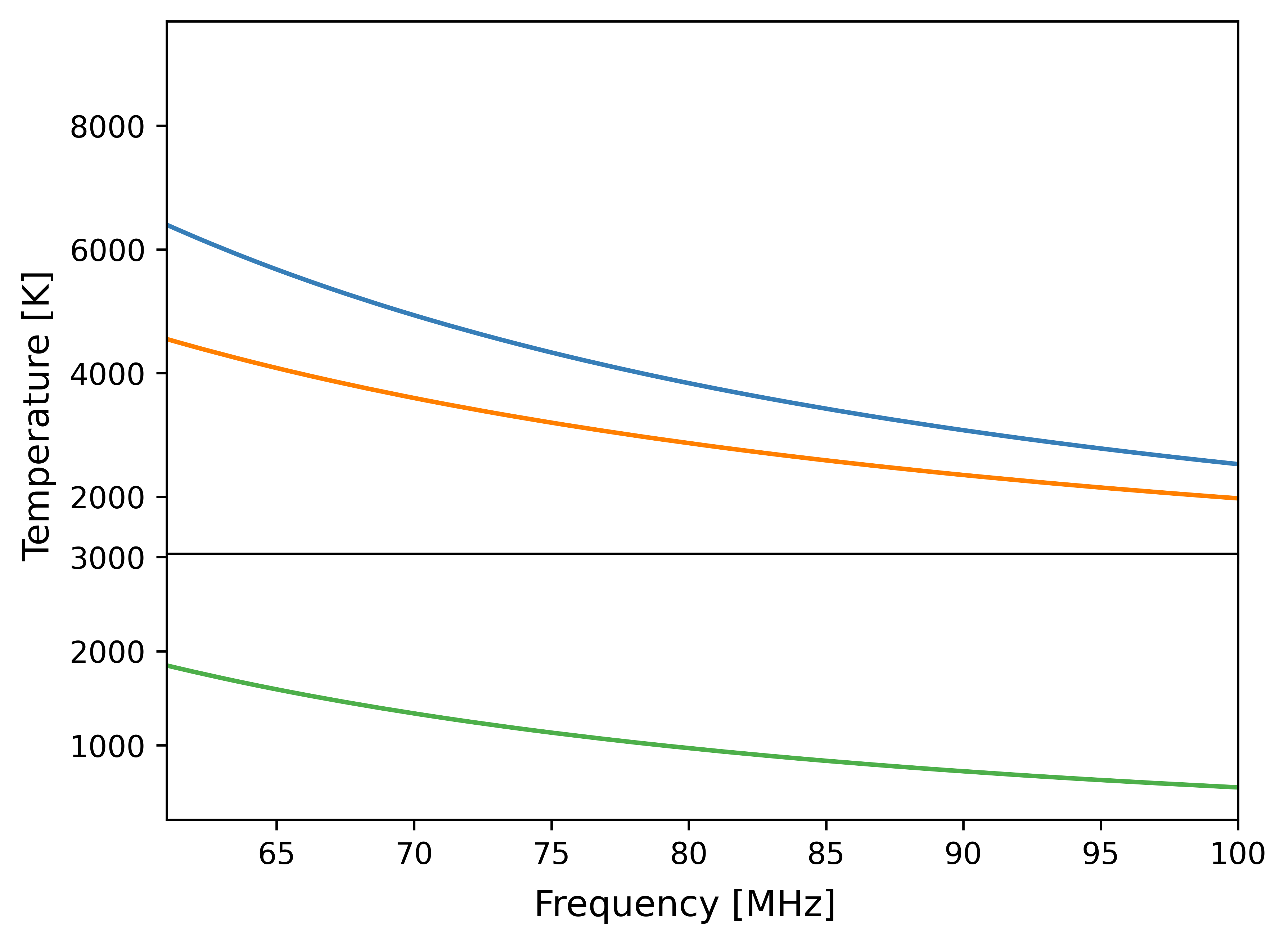}
    \caption{Raw spectrum (dashed blue line) vs calibrated spectrum (solid yellow line) in the antenna simulator laboratory test.  The difference between the two spectra is given below in green.}
    \label{fig:raw_v_cal}
\end{figure}

\begin{figure}
    \centering
    \includegraphics[width=\linewidth]{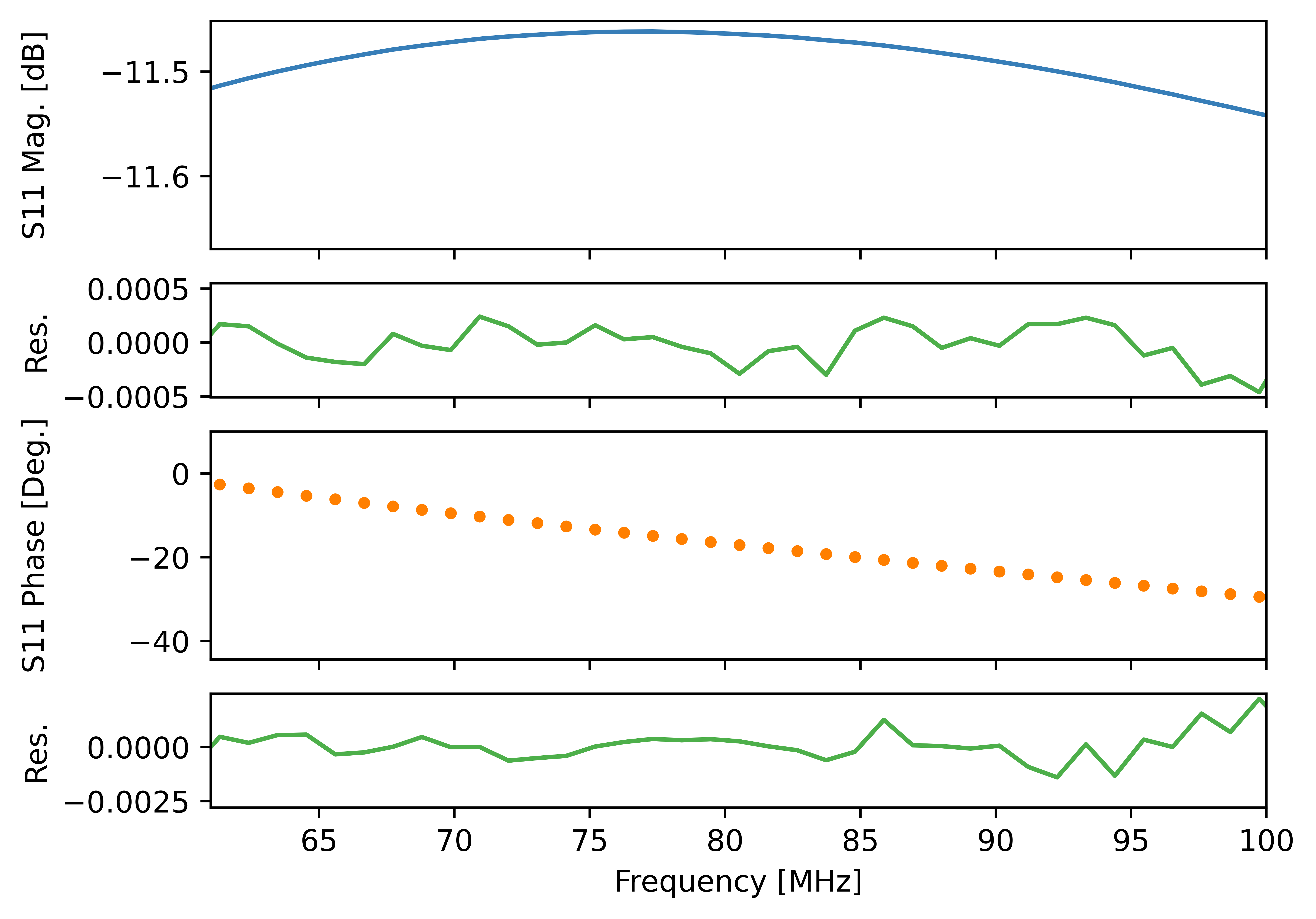}
    \caption{11-term fit to antenna S11, with residuals below in green.}
    \label{fig:ant_S11}
\end{figure}

\begin{figure}
    \centering
    \includegraphics[width=\linewidth]{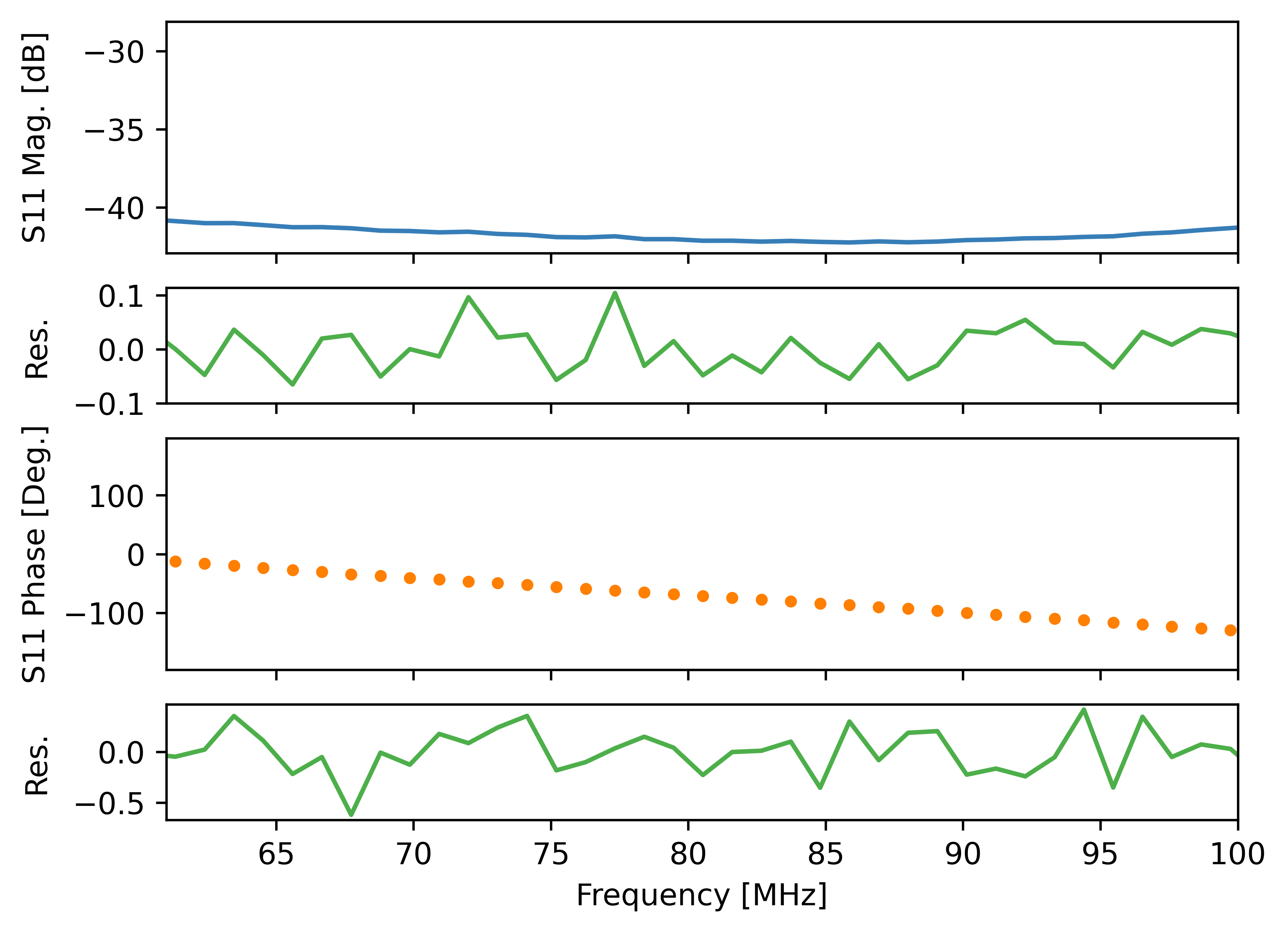}
    \caption{27-term fit to LNA S11, with residuals below in green.}
    \label{fig:lna_S11}
\end{figure}

\begin{figure*}
    \subfigure{\includegraphics[width=0.45\linewidth]{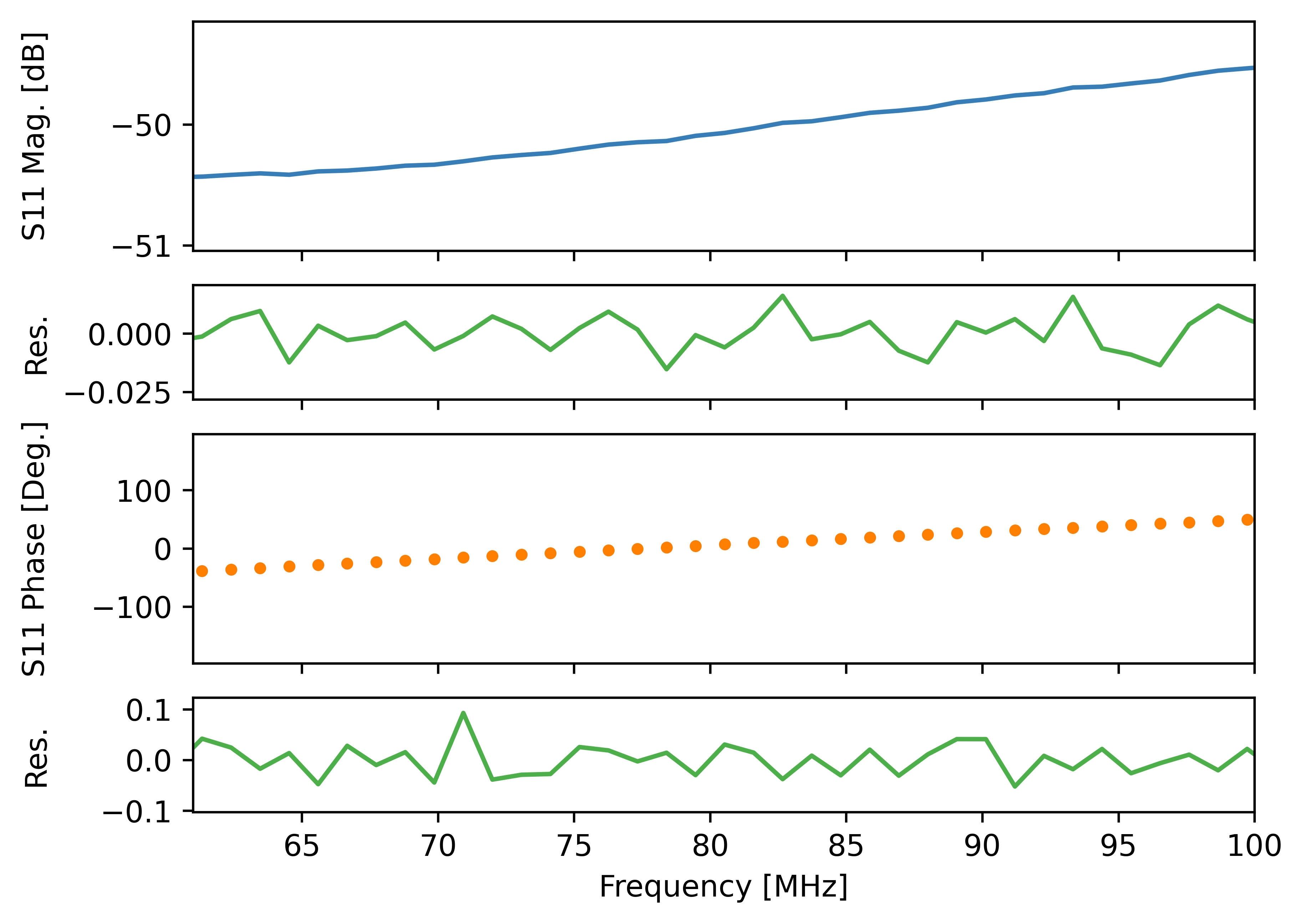}}
    \subfigure{\includegraphics[width=0.45\linewidth]{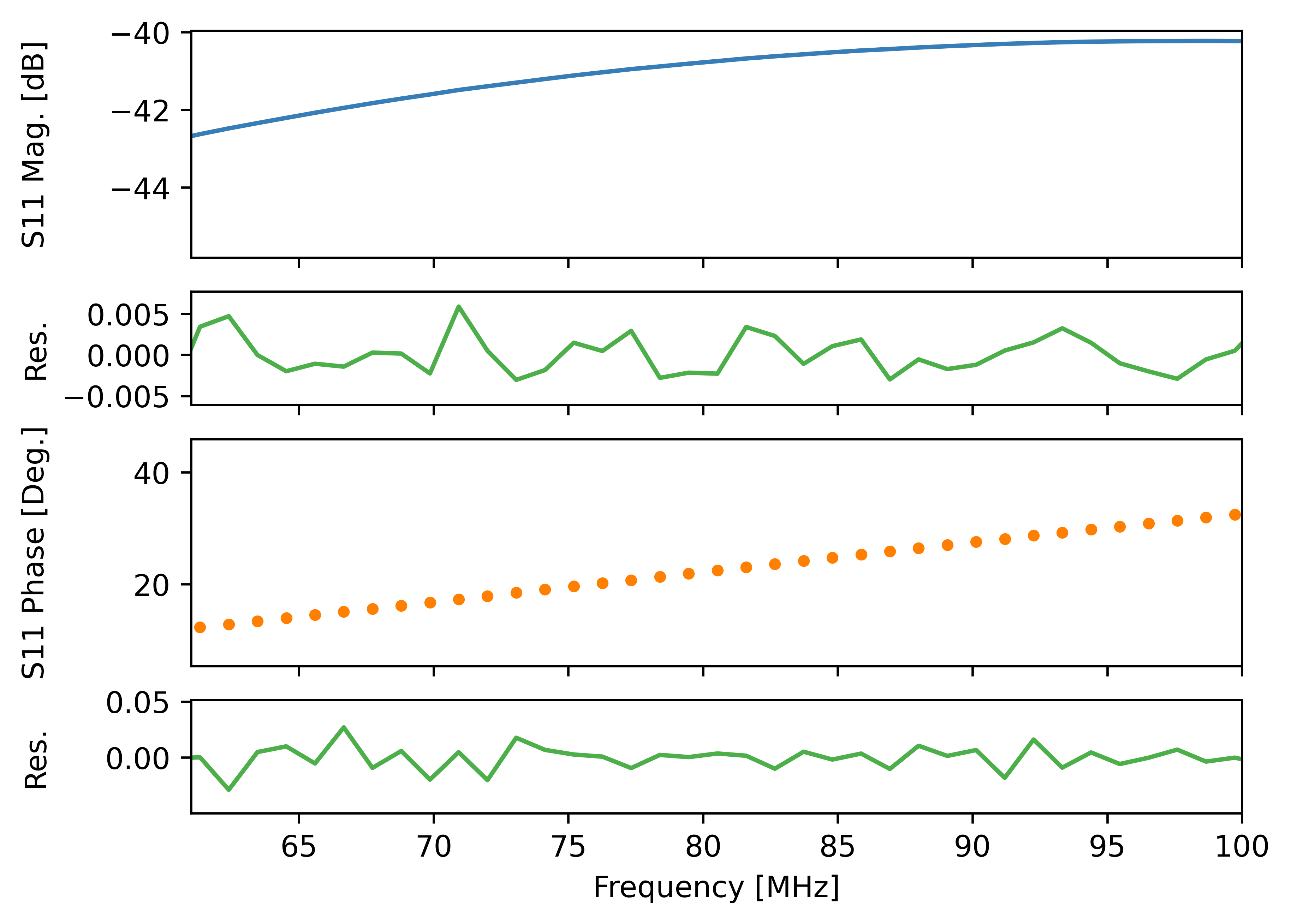}}
    \subfigure{\includegraphics[width=0.45\linewidth]{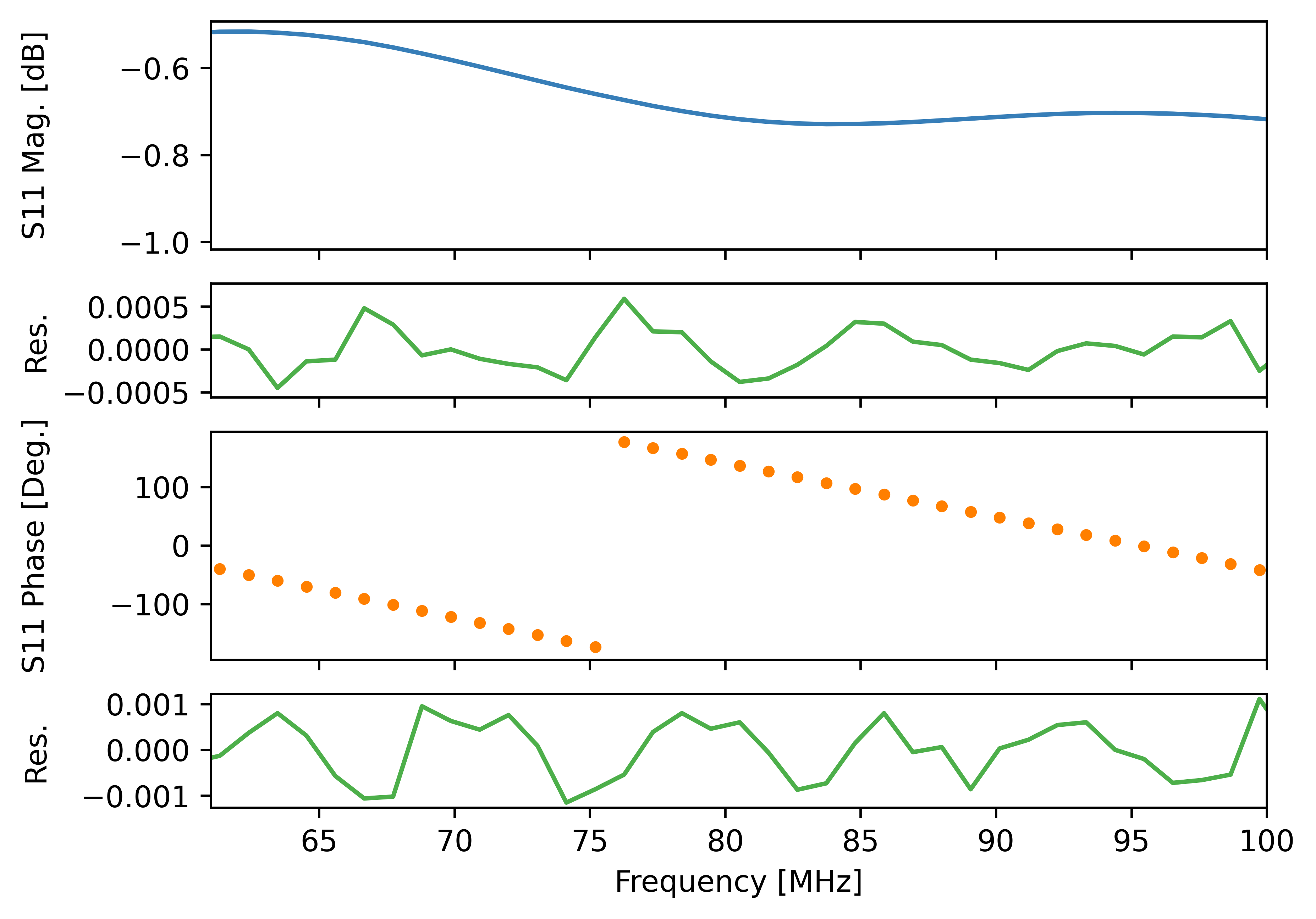}}
    \subfigure{\hspace{17mm} \includegraphics[width=0.45\linewidth]{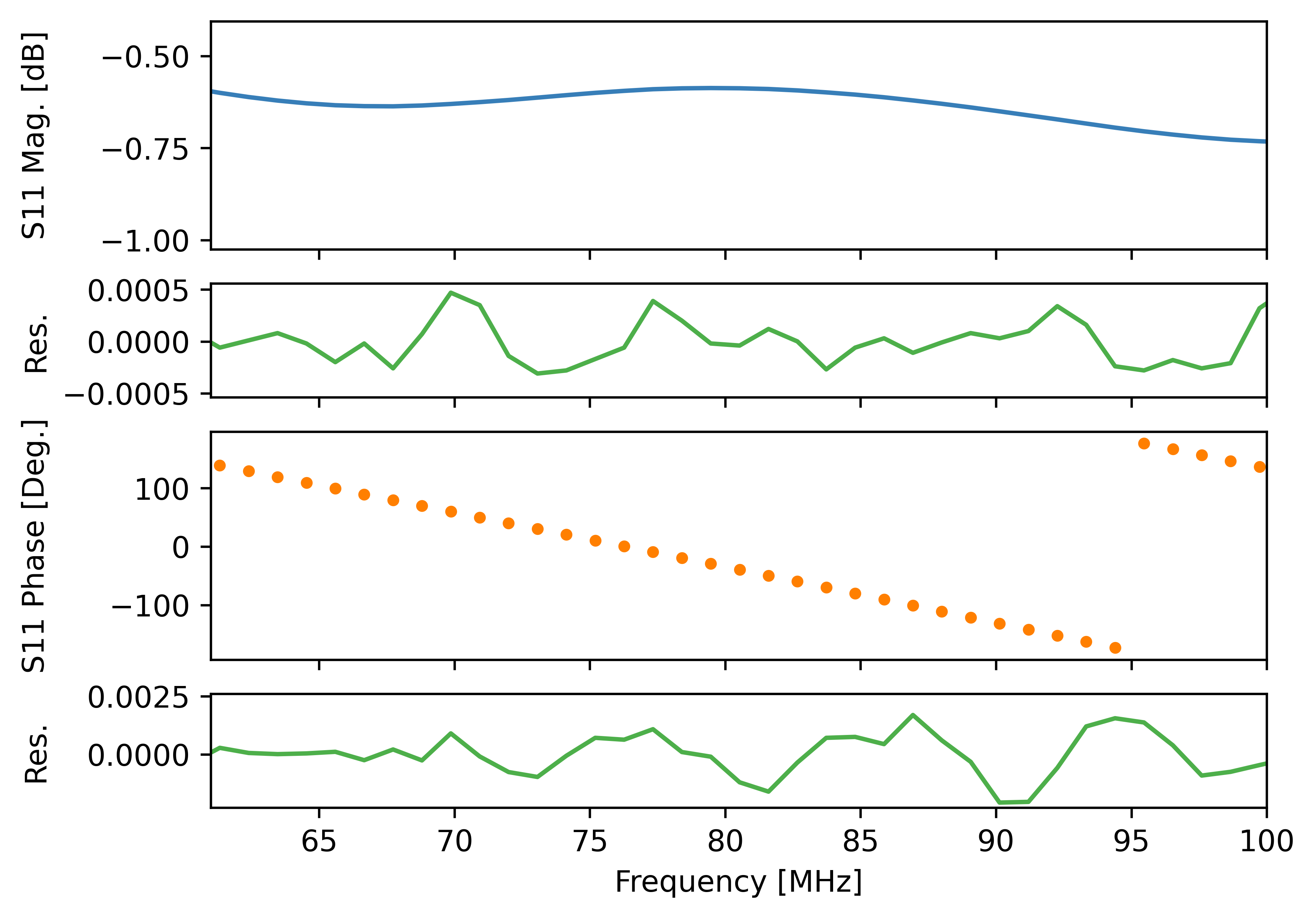}}
    \caption{\textit{Top left:}~27-term fit to the ambient load S11.~\textit{Top right:}~27-term fit to the hot load S11.~\textit{Bottom right:}~27-term fit to the shorted cable S11.~\textit{Bottom left:}~27-term fit to the open cable S11.  All residuals are displayed below in green.}
    \label{fig:hot_amb_open_short}
    \vspace{8mm}
\end{figure*}

\begin{figure}
    \centering
    \includegraphics[width=\linewidth]{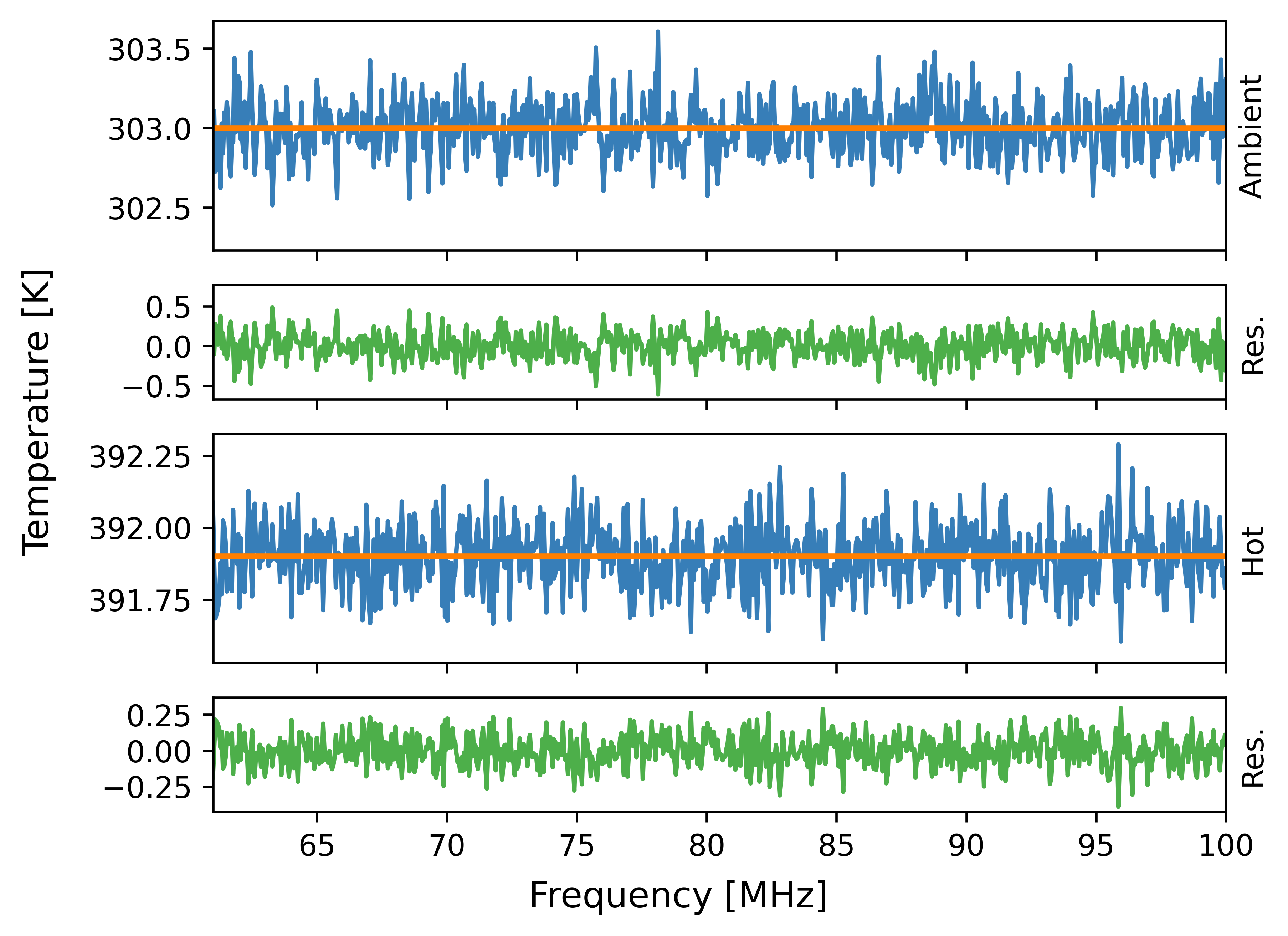}
    \caption{\textit{Top:}~The ambient temperature (303.0 K, yellow line) and the calibrated spectrum of the ambient load (blue).  Residuals are shown below in green, with an RMS of 170 mK.  \textit{Bottom:}~The hot load temperature (391.9 K, yellow line) and the calibrated spectrum of the hot load (blue).  Residuals are shown below in green, with an RMS of 105 mK.}
    \label{fig:amb_hot_cal}
\end{figure}

\begin{figure}
    \centering
    \includegraphics[width=\linewidth]{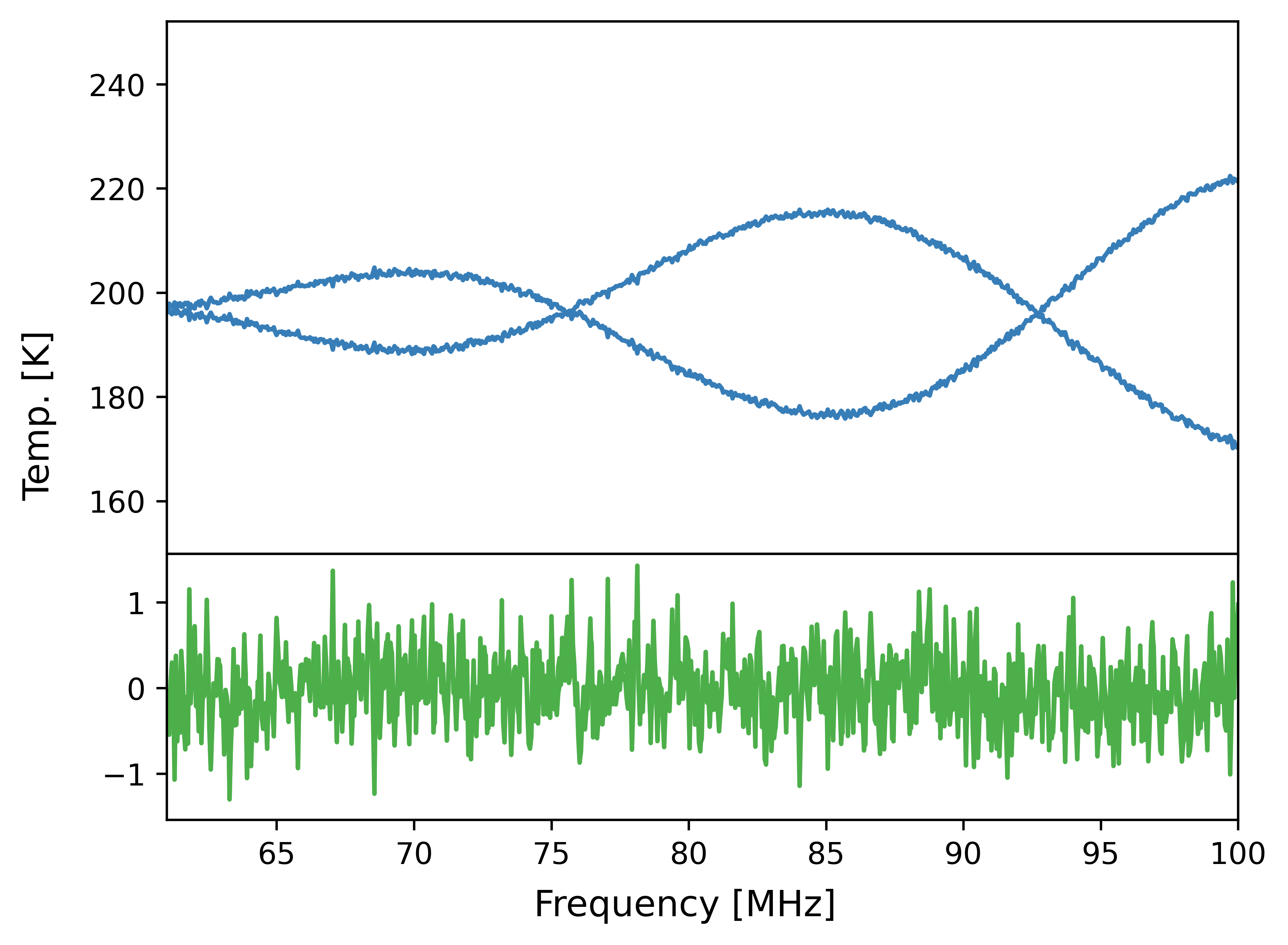}
    \caption{The cable spectrum vs. the model fit to the cable spectrum.  The residuals are given below in green, with an RMS of 413 mK.}
    \label{fig:cable_spec}
\end{figure}

\begin{figure}
    \centering
    \includegraphics[width=\linewidth]{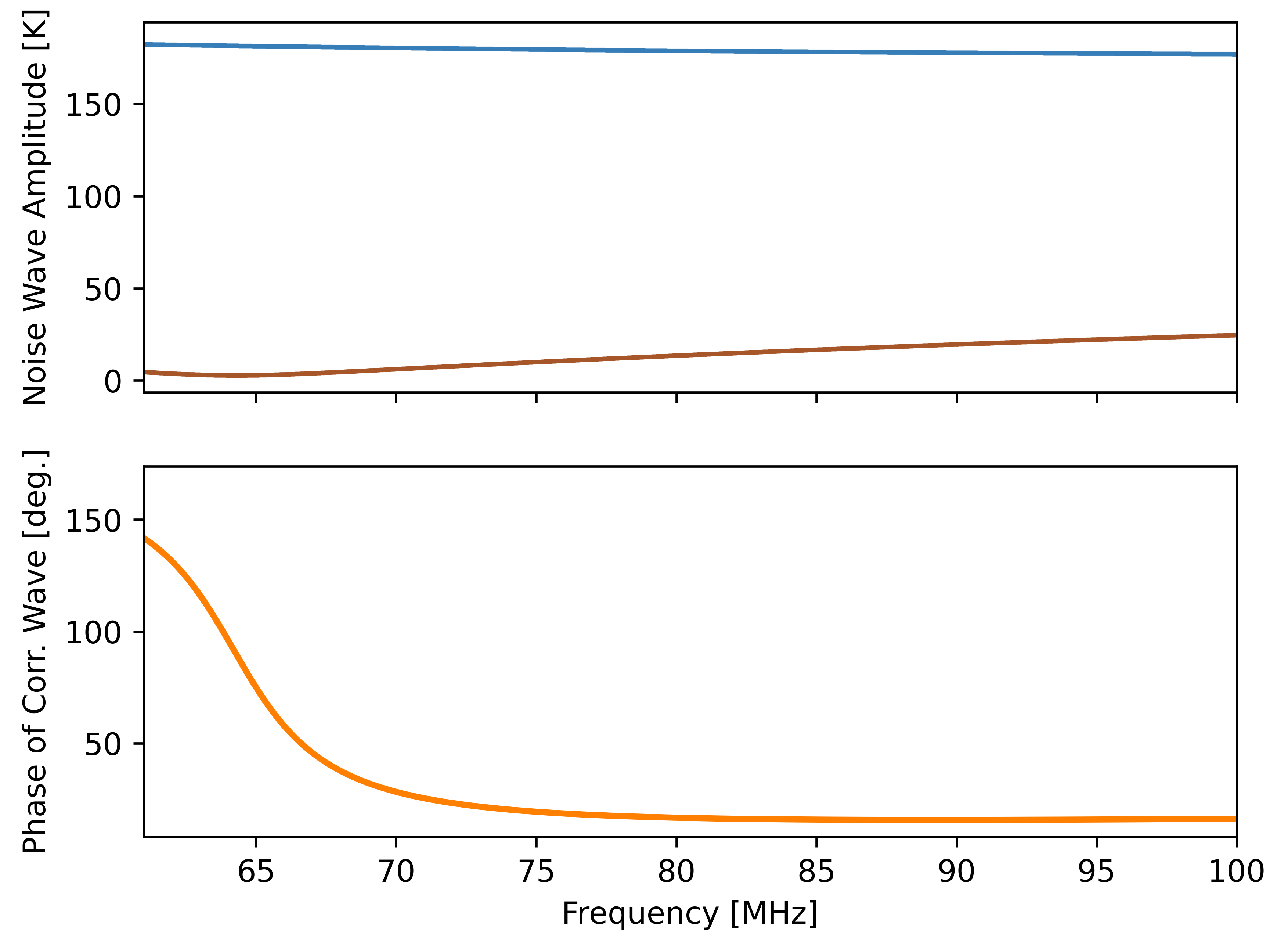}
    \caption{5-term fit to noise waves.  In the top panel, the blue line is the amplitude of the uncorrelated noise, and the brown line is the magnitude of the correlated noise, which is below 50 K throughout the band, and significantly less than 50 K at the lower frequencies.  The bottom panel shows the phase of the correlated wave in degrees.}
    \label{fig:noise_wave}
\end{figure}

Laboratory testing using a source with a known spectrum is requisite prior to deployment.  This is achieved by connecting an antenna simulator to the antenna input.  The results of an initial EDGES-3 antenna simulation test are given in Figures~\ref{fig:raw_v_cal}~through~\ref{fig:noise_wave}.  In this case, the simulated antenna was a noise source combined with a low-pass filter of inductance 0.56 $\mu$H and a 6 dB attenuator to produce a smooth spectrum of $\sim$ 9000 K at 50 MHz, dropping to $\sim$ 2000 K at 120 MHz (Figure \ref{fig:raw_v_cal}), and an antenna simulator S11 of $\sim$~-12 dB (Figure~\ref{fig:ant_S11}).  A high-order log polynomial was fit to each of the calibration spectra, leaving residuals with RMS $\lesssim$ 0.1 dB. 

\begin{figure}
    \centering
    \includegraphics[width=\linewidth]{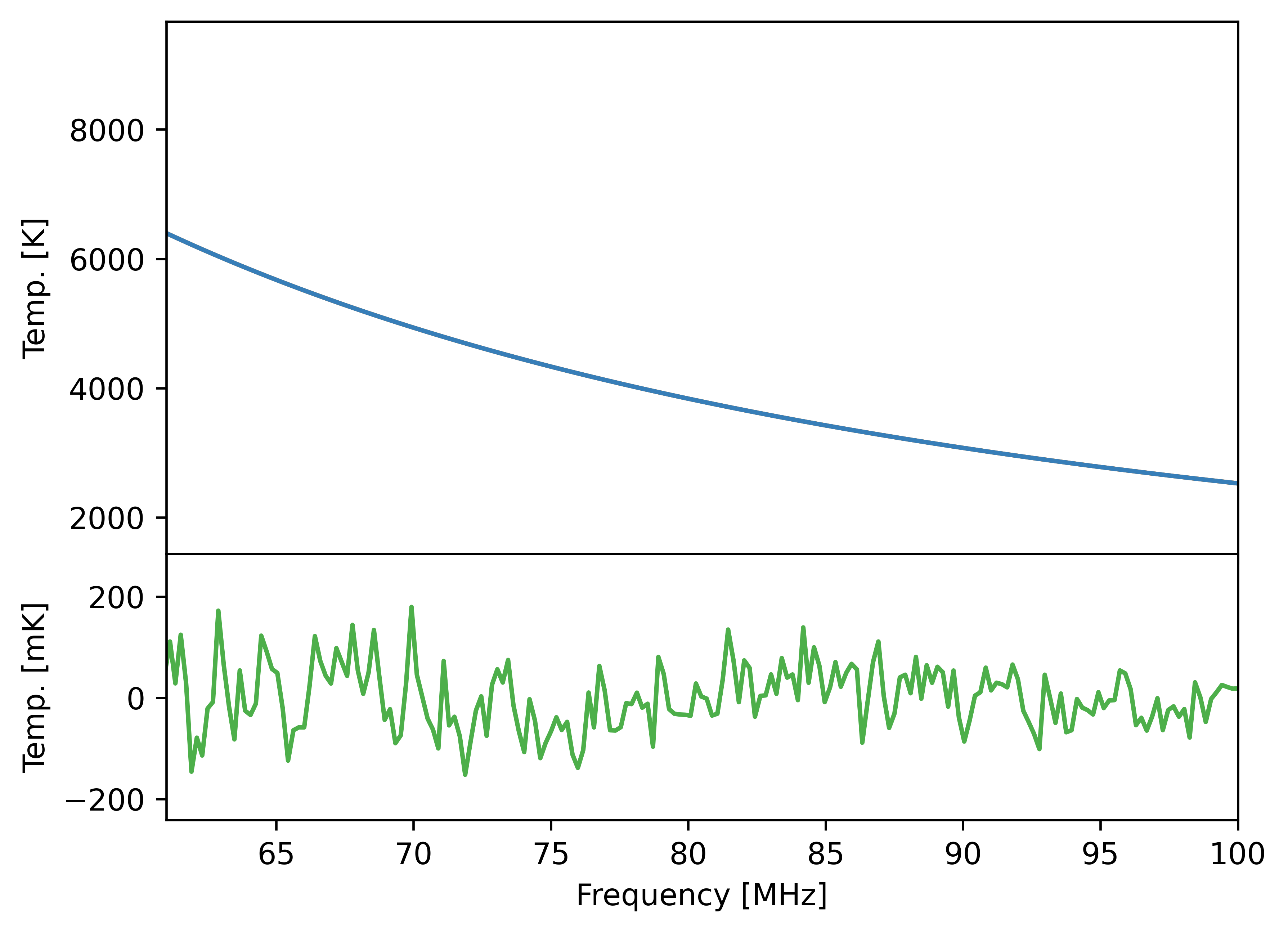}
    \caption{The calibrated, simulated sky spectrum vs a 5-term polynomial fit to the spectrum, with residuals below in green.  The RMS of the resulting residuals is 53 mK.}
    \label{fig:sky_fit}
\end{figure}

The initial system test performed as expected and is documented in HEM \#305 \citep{Memo305}.  The LNA was well-matched to 50~Ohms with an S11 below $-30$~db across the band (Figure \ref{fig:lna_S11}), as was the matched load ambient and hot temperature standard (top left and right panels of Figure \ref{fig:hot_amb_open_short}, respectively).  The matched load standard S11 was below $-45$~dB at the receiver's temperature set point of 25\degree~C and increases to -32~dB when heated to about 127\degree~C.  The long cable reflection coefficients were large (of order 0.5~dB), as expected for both shorted and open terminations with phase slopes matching the expected delay of $\sim26$~ns for a round trip signal (bottom left and right panels of Figure \ref{fig:hot_amb_open_short}, respectively).  The calibrated ambient and hot load noise temperatures were flat across the band to within the noise RMS of $\sim$~100~mK in 6~kHz channels (Figure~\ref{fig:amb_hot_cal}).  The spectra of the long cable (Figure \ref{fig:cable_spec}) and other calibration sources were well fit by the calibration model, all yielding residuals that were flat to the noise RMS of the measurements.  The coefficients of the noise wave parameters \citep{Meys1978} were determined, with the resulting phase and correlated and uncorrelated amplitudes presented in Figure~\ref{fig:noise_wave}.  The resulting calibrated antenna simulator spectrum yielded a nearly-flat spectrum with the residuals from a 5-term polynomial fit showing smooth structure with 53~mK RMS (Figure~\ref{fig:sky_fit}).  

Directly after this laboratory calibration test with an antenna simulator, an absorption feature with identical parameters to the one reported in \cite{Bowman2018} was added via software to the simulated sky spectrum.  Following the same calibration and analysis techniques used in that paper, the test produced a detection of the absorption consistent with the added feature, with an SNR of~12 and a residual RMS of 100~mK over 48~hours of integration.  For more information on this antenna simulation technique and a more comprehensive examination of these results, see HEMs \#305 and \#382 \citep[and ref. therein]{Memo305,Memo382}.

\subsection{Antenna and Ground Plane Characterization}
\label{sec:WA_ground}

Construction of the EDGES-3 ground plane for the permanent installation of the EDGES-3 system at the WA observatory took place in November 2022 (HEM \#406, \citealt{Memo406}).  First light with the full system occurred on November 15$^{\textrm{th}}$, 2022.  The EDGES-3 ground plane and antenna were aligned approximately east-west/north-south with an azimuth of 359\degree~such that the western edge of the ground plane is rotated $\sim$~1\degree~to the north.  The E-plane of the box-blade dipole is parallel to the ground plane along the approximately east-west axis with an azimuth of 269\degree, placing its antenna power pattern nulls in the directions of the nearby electronics hut to the east and the ASKAP dish to the west. 

\begin{figure*}
    \centering
    \includegraphics[width=\linewidth]{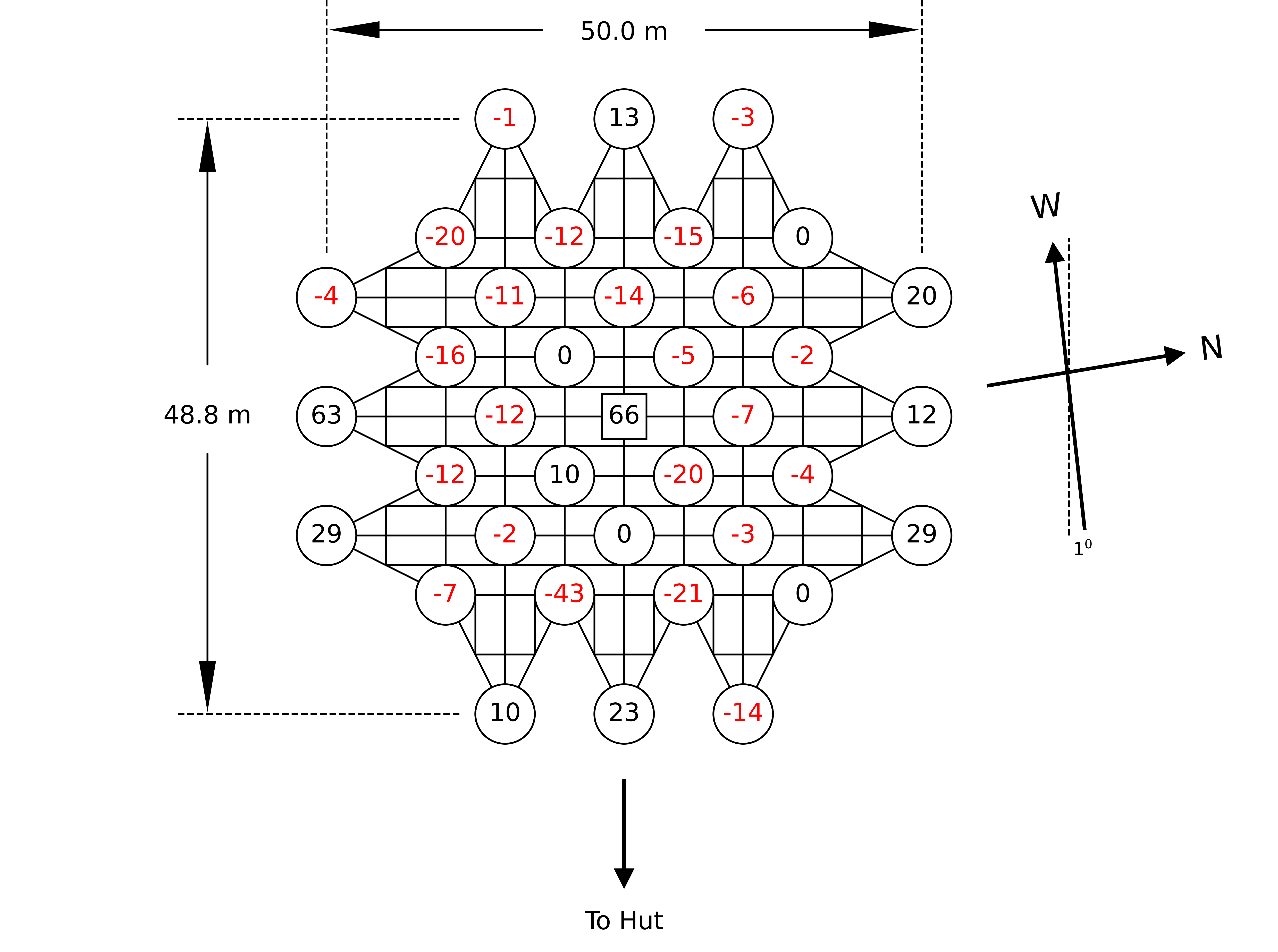}
    \caption{Deviation from planarity in the WA ground plane.  Samples of the ground plane elevation were measured with a laser level immediately after deployment.  Black numbers denote the distance in mm above the best-fit plane, while red numbers are the distance below the best-fit plane, in mm.  Due to the inherent errors in measuring elevation with a laser level and meterstick, the uncertainty on each of these values is $\pm$~8 mm.  There is a gentle slope over the ground plane footprint, rising a maximum of $\sim$~0.3 m from NW to SE.  This rise occurs over $\sim$~50 m, suggesting a maximum grade of $\sim$~0.7~\%.  Note that the base plate in the center is significantly higher than the surrounding ground plane, as addressed in (\S~\ref{sec:welded}).}
    \label{fig:EDGES3-ground-plane-planarity}
    \vspace{10mm}
\end{figure*}

The soil surface under EDGES-3 is naturally planar and reasonably level with a shallow slope down from south to north ($\sim100$~mm tip to tip) and west to east ($\sim300$~mm tip to tip).  The base structure rests on the soil and is level to within 0.07\degree~along the east-west axis and slopes slightly down from north to south by 0.9\degree.  The wire mesh sheets of the ground plane rest directly on the soil.  Some wire mesh sheets were slightly bent during transportation and assembly.  We assessed deviations in the ground plane planarity immediately after assembly.  Figure~\ref{fig:EDGES3-ground-plane-planarity} shows deviations in height relative to the best-fit plane at a grid of points across the mesh.  The RMS of the deviations is 21~mm.  These deviations vary smoothly over scales of order a meter or two across the ground plane.  Electromagnetic modeling for EDGES-3 indicates such non-planarity does not significantly affect the observations as long as the amplitude of the deviations are $\lesssim$~50~mm (HEM \#383, \citealt{Memo383}).  We have observed on previous EDGES deployments that similar variations eventually diminish over 1-2 years, likely as the metal of the mesh relaxes to the soil surface.

\subsubsection{Resonance at 75~MHz}
\label{sec:75_res}

During early operation of the EDGES-3 at the WA observatory there was unexpectedly large spectral structure centered around 75~MHz discovered after performing low-order polynomial fits and subtraction to observations.  This structure varied in amplitude with LST, which is characteristic of a resonance since the strength of the signal scales with the sky noise power, which changes as the galactic center passes overhead.  Through modeling of possible causes, we identified the spacing of the bolts connecting the aluminum panels on the base structure to its steel frame as the likely cause, as described in HEM \#408 \citep{Memo408}.  For the 150~cm bolt spacing, small gaps\footnote{Widths of $\sim$~0.3 mm lead to an equivalent 40 pf capacitance load at the center of the slot.} between the panels and steel frame in the sections between bolts created slot antennas with resonances in the EDGES-3 band.  On February 23$^{\textrm{rd}}$, 2023, additional bolts were added to the base plate to reduce the spacing to 75~mm, moving any such resonance out of the primary EDGES-3 band.  Subsequent observations showed the resonance was eliminated, or at least diminished to the extent that it could no longer be distinguished from noise. 

\subsubsection{Mechanical Stability of the Antenna}

During initial operations, it was found the antenna boxes were not maintaining a constant separation (originally 36~mm), but were instead separating as much as 37.9~mm over diurnal cycles and longer.  Custom 3D-printed plastic spacers were added between the antenna boxes to provide a rigid separation guide at 35~mm.  This modification was completed at the end of May, 2023, and the instrument has operated nominally since then.

\subsubsection{Field sensitivities}

\begin{figure}
    \centering
    \includegraphics[width=\linewidth]{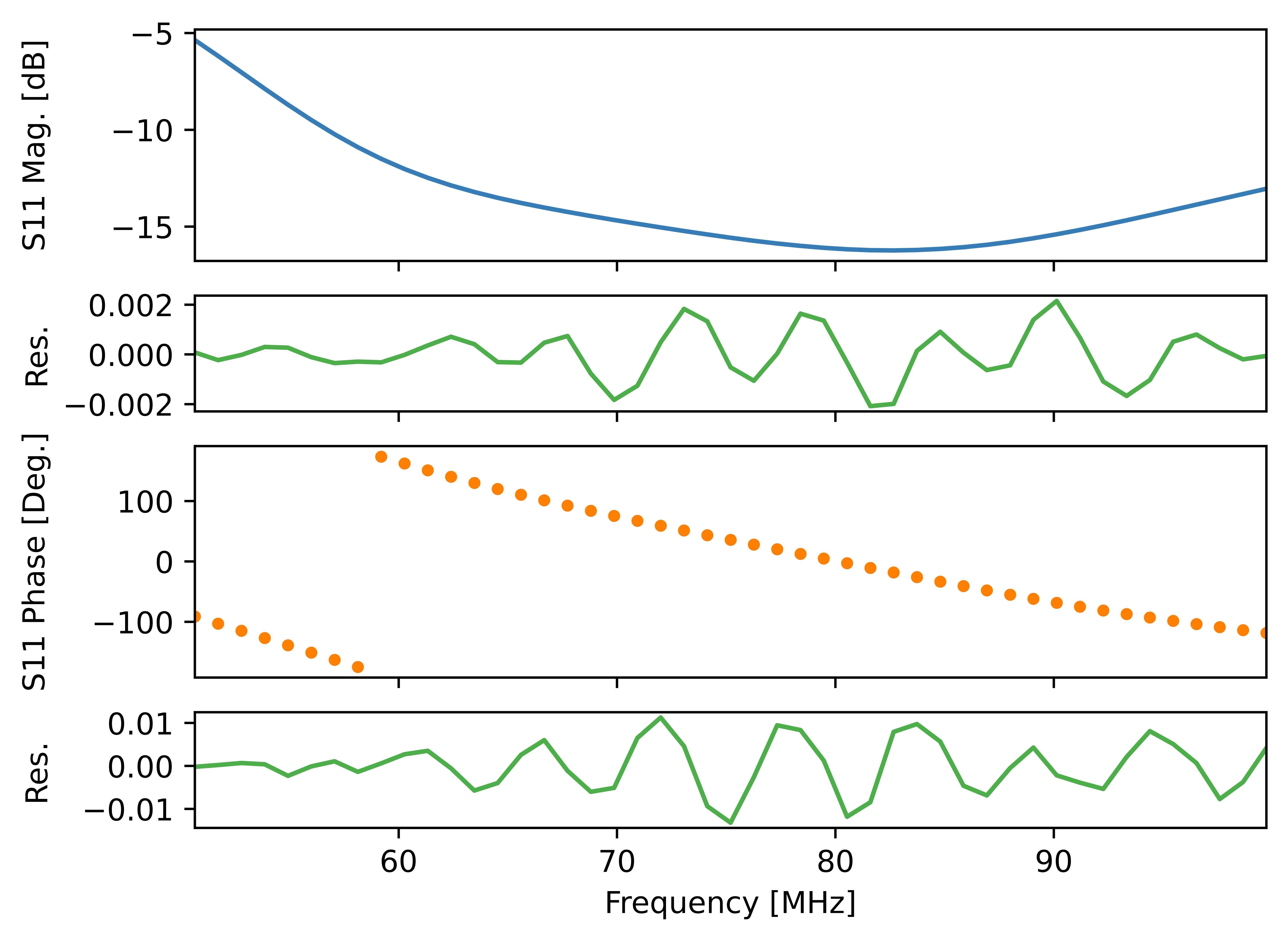}
    \caption{11-term fit to the antenna S11 in the field at the WA Observatory, taken March 11$^{\textrm{th}}$, 2023.}
    \label{fig:field_ant_S11}
\end{figure}

Sensitivities in the fully deployed EDGES-3 system to various systematics were calculated once nominal operation was achieved, and are presented in Table \ref{table:obslog}.  These sensitivities were calculated after removing 5 physical terms using the 408 MHz Haslam sky map \citep{Haslam1995}, scaled down to 50 - 100 MHz with a -2.5 spectral index.  An example of an antenna S11 from the field at WA Observatory is presented in Figure~\ref{fig:field_ant_S11}.

\begin{table}[ht!]
\centering
\caption{Sensitivities of EDGES-3 at the WA observatory} 
\label{tab:sense}
\begin{tabular}{ll}
\hline
Parameter & Figure of Merit             \\ 
\hline 
VNA fractional error & below 0.0003 \\
LNA uncorrelated noise & 180 K \\
LNA correlated noise & below 30 K \\
LNA input reflection coefficient & below -39 dB \\
Antenna reflection coefficient & below -10 dB \\
Antenna RMS beam chromaticity\textbullet\dag & 67 mK \\
Antenna RMS beam chromaticity\textbullet\ddag & 4 mK \\
Antenna + ground plane loss & below 0.3\% \\
Antenna gain at 1$^\textrm{o}$ elevation & below -25 dBi \\
\hline
\end{tabular} \\
\vspace{2mm}
\textbullet~5 physical terms removed using 408 MHz Haslam sky map scaled with -2.5 spectral index to 50 - 100 MHz \\
\dag averaged over 1 hour blocks of LST \\
\ddag averaged over 24 hour blocks of LST \\

\label{table:obslog} 
\end{table}

\section{Conclusions} \label{sec:conc}

This work examines multiple facets of a single-antenna experiment designed to observe the global 21-cm absorption signal produced during cosmic dawn.  First, it addresses the challenges inherent in designing an experiment capable of detecting such a weak signal on top of a strong background (\S~\ref{sec:challenges}).  Then, it describes the design of EDGES-3, and how various improvements and additions to the EDGES-2 system were incorporated specifically to address these myriad challenges (\S~\ref{sec:edges3}).  For clarity, the primary takeaways from these two sections are distilled in Table \ref{tab:mitigation}.  Finally, the long-term deployment of an EDGES-3 system in WA is described, with testing and commissioning data presented (\S~\ref{sec:commissioning}).  Shortly, a data analysis paper will be released, employing a newly-developed analysis pipeline to further probe the validity of the EDGES-2 detection announced in \cite{Bowman2018}.


\section*{Acknowledgments}

This research was funded by the following National Science Foundation grants: AST-1609450, AST-1813850, and AST-1908933.  
This project has also received funding from the European Union’s Horizon 2020 research and innovation programme under the Marie Skłodowska-Curie grant agreement No 101067043.  EDGES is located at the Inyarrimanha Ilgari Bundara, the CSIRO Murchison Radio-astronomy Observatory.  We acknowledge the Wajarri Yamatji people as the traditional owners of the Observatory site.  We also thank CSIRO for providing site infrastructure and on-going support.

\FloatBarrier

\bibliography{References}{}
\bibliographystyle{aasjournal}



\end{document}